\documentclass[ twocolumn]{aastex631}

\newcommand{\cii}{C\,\textsc{II}}
\newcommand{\oii}{[O\,\textsc{II}]}
\newcommand{\oiii}{[O\,\textsc{III}]}
\newcommand{\nii}{[N\,\textsc{II}]}

\newcommand{\feiii}{[Fe\,\textsc{III}]}

\newcommand{\cliii}{[Cl\,\textsc{III}]}
\newcommand{\siii}{[S\,\textsc{III}]}
\newcommand{\sii}{[S\,\textsc{II}]}
\newcommand{\ariv}{[Ar\,\textsc{IV}]}
\newcommand{\hi}{H\,\textsc{I}}
\newcommand{\hii}{H\,\textsc{II}}

\newcommand{\hei}{He\,\textsc{I}}

\newcommand{\oiirls}{O\,\textsc{II}}

\usepackage{amsmath}

\begin{document}

\title{Generalized $T_{\rm e}$(\oiii)$-$$T_{\rm e}$(\hei) Discrepancies in Ionized Nebulae: Possible Evidence of Case B Deviations and Temperature Inhomogeneities}

\author[0000-0002-6972-6411]{J. E. M\'endez-Delgado}
\affiliation{Instituto de Astronom\'ia, Universidad Nacional Aut\'onoma de M\'exico, Ap. 70-264, 04510 CDMX, M\'exico}
\affiliation{Astronomisches Rechen-Institut, Zentrum f\"ur Astronomie der Universit\"at Heidelberg, Mönchhofstraße 12-14, D-69120 Heidelberg, Germany}
\correspondingauthor{J. Eduardo M\'endez-Delgado}
\email{jmendez@astro.unam.mx}

\author[0000-0003-0605-8732]{E. D. Skillman}
\affiliation{Minnesota Institute for Astrophysics, University of Minnesota, 116 Church Street South East, Minneapolis, MN 55455, USA}

\author{E. Aver}
\affiliation{Department of Physics, Gonzaga University, 502 E Boone Ave., Spokane, WA 99258, USA}

\author[0000-0001-5801-6724]{C. Morisset}
\affiliation{Instituto de Astronom\'ia, Universidad Nacional Aut\'onoma de M\'exico, Ap. 106, 22800 Ensenada, Baja California, M\'exico}
\affiliation{Instituto de Ciencias F\'isicas, Universidad Nacional Aut\'onoma de M\'exico, Av. Universidad s/n, 62210 Cuernavaca, Mor., M\'exico}

\author[0000-0002-5247-5943]{C. Esteban}
\affiliation{Instituto de Astrof\'isica de Canarias, E-38205 La Laguna, Tenerife, Spain}
\affiliation{Departamento de Astrof\'isica, Universidad de La Laguna, E-38206 La Laguna, Tenerife, Spain}

\author[0000-0002-6138-1869]{J. Garc\'ia-Rojas}
\affiliation{Instituto de Astrof\'isica de Canarias, E-38205 La Laguna, Tenerife, Spain}
\affiliation{Departamento de Astrof\'isica, Universidad de La Laguna, E-38206 La Laguna, Tenerife, Spain}

\author[0000-0001-6551-3091]{K. Kreckel}
\affiliation{Astronomisches Rechen-Institut, Zentrum f\"ur Astronomie der Universit\"at Heidelberg, Mönchhofstraße 12-14, D-69120 Heidelberg, Germany}

\author[0000-0002-0361-8223]{N. S. J. Rogers}
\affiliation{Department of Physics and Astronomy, Northwestern University, 2145 Sheridan Road, Evanston, IL 60208, USA}
\affiliation{Center for Interdisciplinary Exploration and Research in Astrophysics (CIERA), Northwestern University, 1800 Sherman Avenue, Evanston, IL 60201, USA}

\author[0000-0002-3642-9146]{F. F. Rosales-Ortega}
\affiliation{Instituto Nacional de Astrof\'isica, \'Optica y Electr\'onica (INAOE-CONAHCyT), Luis E. Erro 1, 72840, Tonantzintla, Puebla, M\'exico}

\author[0000-0002-2644-3518]{K. Z. Arellano-C\'ordova}
\affiliation{Institute for Astronomy, University of Edinburgh, Royal Observatory, Edinburgh, EH9 3HJ, United Kingdom}

\author[0000-0002-0159-2613]{S. R. Flury}
\affiliation{Institute for Astronomy, University of Edinburgh, Royal Observatory, Edinburgh, EH9 3HJ, United Kingdom}

\author[0000-0003-1192-6987]{E. Reyes-Rodr\'iguez}
\affiliation{Departamento de Astrof\'isica, Universidad de La Laguna, E-38206 La Laguna, Tenerife, Spain}
\affiliation{Isaac Newton Group of Telescopes, Apto 321, E-38700 Santa Cruz de La Palma, Canary Islands, Spain}

\author[0000-0002-0539-1720]{M. Orte-Garc\'ia}
\affiliation{Instituto de Astrof\'isica de Canarias, E-38205 La Laguna, Tenerife, Spain}
\affiliation{Departamento de Astrof\'isica, Universidad de La Laguna, E-38206 La Laguna, Tenerife, Spain}

\author[0009-0006-7127-2857]{S. Tan}
\affiliation{Institut f\"ur Theoretische Astrophysik, Zentrum f\"ur Astronomie der Universit\"at Heidelberg, Albert-Ueberle-Str 2, D-69120 Heidelberg, Germany}



\begin{abstract}

The physics of recombination lines (RLs) in the \hei~singlet system is expected to be relatively simple, supported by accurate atomic models. We examine the intensities of \hei~singlets $\lambda \lambda 3614, 3965, 5016, 6678, 7281$ and the triplet \hei~$\lambda 5876$ in various types of ionized nebulae and compare them with theoretical predictions to test the validity of the ``Case B'' recombination scenario and the assumption of thermal homogeneity. Our analysis includes 85 spectra from Galactic and extragalactic \hii~regions, 90 from star-forming galaxies, and 218 planetary nebulae, all compiled by the DEep Spectra of Ionized REgions Database Extended (DESIRED-E) project. By evaluating the ratios \hei~$\lambda 7281/\lambda 6678$ and \hei~$\lambda 7281/\lambda 5876$, we determine $T_{\rm e}$(\hei) and compare it with direct measurements of $T_{\rm e}$(\oiii~$\lambda 4363/\lambda 5007$). We find that $T_{\rm e}$(\hei) is systematically lower than $T_{\rm e}$(\oiii) across most objects and nebula types. Additionally, we identify a correlation between the abundance discrepancy factor (ADF(O$^{2+}$)) and the difference $T_{\rm e}$(\oiii) - $T_{\rm e}$(\hei) for planetary nebulae. We explore two potential explanations: photon loss from $n^1P \rightarrow 1^1S$ transitions and temperature inhomogeneities. Deviations from ``Case B'' may indicate photon absorption by \hi~rather than \hei~and/or generalized ionizing photon escape, highlighting the need for detailed consideration of radiative transfer effects. If temperature inhomogeneities are widespread, identifying a common physical phenomenon affecting all ionized nebulae is crucial. Our results suggest that both scenarios can contribute to the observed discrepancies.

\end{abstract}



\section{Introduction}
\label{sec:intro}

Helium is the second most abundant element in the universe, and the precise determination of its abundance is of profound importance for cosmology, as it was created in large quantities during the Big Bang \citep{Alpher:48, Peebles:66, Peebles:66b, Peimbert:70, Peimbert:74}. In star-forming regions, \hei~lines are frequently detected in a large number of spectra. These are recombination lines (RLs), produced after the capture of free electrons by ionized helium atoms. Since \hei~is an atomic system with two electrons, this atom has two level states depending on its total spin quantum number: singlets, and triplets \citep{Heisenberg:26}.

Approximately three-quarters of the recombinations will occur in the triplet system \citep{Burgess:60}, giving rise to the brightest \hei~lines, such as $\lambda \lambda 4471, 5875, 7065$. However, in the triplet configuration, the $2^3S$ level is metastable, making it susceptible to the effects of self-absorption and collisional excitation and de-excitation. Consequently, the fluxes of the triplet \hei~lines can significantly diverge from pure recombination predictions. For instance, lines such as \hei~$\lambda 7065$ increase their flux at the expense of lines such as \hei~$\lambda 3889$, although the total flux of the triplet system lines must be conserved \citep{Porter:07}. These effects are well known and are considered in great detail in many works dedicated to the precise determination of the fraction of primordial mass in helium $Y_p$ \citep{Izotov:98, Peimbert:02, Aver:15, Valerdi:19}.

In contrast to the triplet system, the physics involved in the emission of singlet lines is relatively simple. The effects of self-absorption and collisional excitation are expected to be very small, and the fluxes of these lines should be well described in terms of pure recombination \citep{Porter:07, Porter:09}. Additionally, the recombination probabilities and atomic data for \hei~are extremely precise \citep{Benjamin:99, Porter:05, Porter:12, Porter:13, DelZanna:22}, with potential errors on the order of $\sim$ 1\%. Therefore, the study of these lines is extremely relevant for testing our most basic (and general) physical assumptions adopted for determining chemical abundances in ionized nebulae, such as the thermal homogeneity \citep[see the discussion in][]{Ferland:16} and the assumption that the nebula is optically thick to Lyman transitions also referred as ``Case B'' \citep{Baker:38}. 

\begin{figure}[h]
\epsscale{1.15}
\plotone{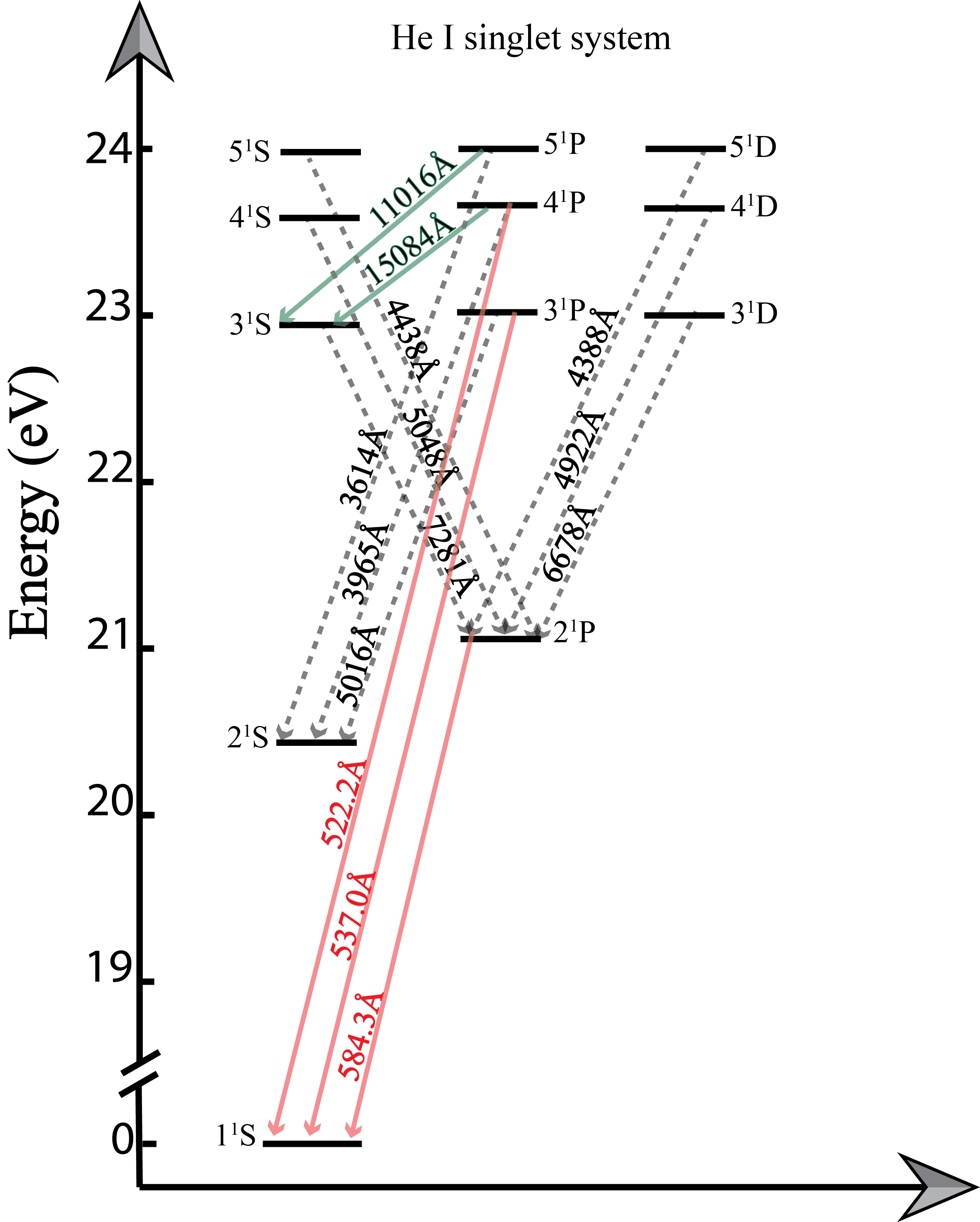}
\caption{Grotrian diagram \citep{Grotrian:28} for the singlet system of \hei. Adapted and extended from fig.~D4 of \citet{MendezDelgado:21a}. Some permitted transitions from the levels $n^1P \rightarrow 1^1S$ are marked in red. The transitions $4^1P \rightarrow 3^1S$ and $5^1P \rightarrow 3^1S$, which give rise to \hei~$\lambda \lambda15084, 11016$ lines, respectively, are the most important for increasing the population of the $3^1S$ level after direct recombination and are highlighted in green.}
\label{fig:grotrian}
\end{figure}

Intensity ratios of \hei~singlet lines originating from $n^1S\rightarrow n^1P$ and $n^1P\rightarrow n^1S$ transitions (such as \hei~$\lambda 7281$ and \hei~$\lambda 5016$, respectively) compared to those originating from $n^1D\rightarrow n^1P$ transitions (such as \hei~$\lambda 6678$) are sensitive to the electron temperature of the gas ($T_{\rm e}$), with a small dependence on the density ($n_{\rm e}$) \citep{Zhang:05}. Additionally, these same ratios are very sensitive to deviations from ``Case B'' since the transitions between the $n^1P$ levels and the ground level $1^1S$ are permitted by the electric-dipole rules (See Fig.~\ref{fig:grotrian}).

When comparing $T_{\rm e}$(\hei~$\lambda 7281$/$\lambda 6678$) with classical nebular diagnostics based on collisionally excited lines (CELs) of high ionization degree, such as $T_{\rm e}$(\oiii~$\lambda 4363/\lambda 5007$), one would expect relatively good consistency, as assumed in multiple studies \citep{Izotov:14, Valerdi:21b, Dors:24}, in consistency with the predictions of photoionization models of star-forming regions (see Sec.~\ref{sec:appendix_temps}). However, in the presence of small-scale temperature variations, $T_{\rm e}$(\oiii) would have a systematic bias toward higher temperatures, while $T_{\rm e}$(\hei) would remain relatively unaffected \citep{Peimbert:67}. Under these conditions, one would observe that $T_{\rm e}$(\hei) $<$ $T_{\rm e}$(\oiii). The presence of these temperature inhomogeneities would imply that most nebular chemical abundance determinations based on CELs would be systematically underestimated and are part of the academic debate due to their profound implications \citep{GarciaRojas:07b, Esteban:09, Cameron:23, MendezDelgado:23a, MendezDelgado:24, Chen:23}.

If deviations from ``Case B'' such as photon loss originating from \hei~$n^1P\rightarrow 1^1 S$ transitions exist, one would see a decrease in the overall flux of \hei~singlets, primarily affecting the lines originating from $n^1P$ levels, followed by those from $n^1S$ and $n^1D$ levels, in that order, due to the permitted interconnection transitions. In that case, one would observe that $T_{\rm e}$(\hei~$\lambda 7281$/$\lambda 6678$) $<$ $T_{\rm e}$(\oiii~$\lambda 4363/\lambda 5007$). If fluorescent excitations \hei~$1^1 S \rightarrow n^1P$ in an optically thick nebula exist \citep[known as ``Case D''-recombination see][]{Luridiana:09}, one would observe the opposite effect, obtaining $T_{\rm e}$(\hei) $>$ $T_{\rm e}$(\oiii). If significant deviations from ``Case B'' are observed in the singlet \hei~atom, this may indicate the widespread presence of fluorescent effects in the radiative transfer of different ions and/or the escape of these ionizing photons into the interstellar medium (ISM).

The main drawback of studying \hei~singlet lines is that they are relatively faint, with the exception of \hei~$\lambda 6678$. Through the DESIRED project \citep{MendezDelgado:23b}, and its extension \citep[DESIRED-E,][]{MendezDelgado:24b}, we have collected an extensive set of deep spectroscopic data from numerous ionized nebulae. This has enabled us to study the physical and chemical properties of nebulae through a homogeneous analysis of very weak emission lines. As a result, we are opening a new frontier in astrophysics by facilitating a comprehensive study of the physical phenomena associated with these faint lines. In this work, we consistently and homogeneously analyze direct determinations of $T_{\rm e}$(\hei) in 175 deep spectra of Galactic and extragalactic star-forming regions. Additionally, we adopt 218  deep spectra of Galactic and extragalactic planetary nebulae with the same criteria as for the star-forming regions to contrast the results under different photoionization conditions. This is the largest sample of ionized nebulae used for this purpose in the literature. In each case, we have compared $T_{\rm e}$(\hei) with $T_{\rm e}$(\oiii).

In Sec.~\ref{sec:obs}, we present the observations adopted in this work, as well as the methodology used to determine the physical conditions and ionic abundances of the gas. In Sec.~\ref{sec:res}, we present the observed trend between $T_{\rm e}$(\hei) and $T_{\rm e}$(\oiii) determined for the observational sample. In Sec.~\ref{sec:disc}, we discuss the results in several subsections, considering various phenomena and models that could explain the unexpected trend observed in Sec.~\ref{sec:res}. Finally, in Sec.~\ref{sec:concs}, we summarize our conclusions and final thoughts. In Appendix~\ref{sec:appendix}, we present the relevant data, calculations, and observational values for this study, as well as additional figures that complement our findings.

\section{Observations and Methodology}
\label{sec:obs}

The observational sample used for this study comes from the deepest optical spectra published to date, collected and treated homogeneously as part of the DESIRED and DESIRED-E projects \citep{MendezDelgado:23b,MendezDelgado:24b}. In these projects, the reddening-corrected fluxes of all emission lines reported by the referenced authors have been compiled, surpassing previous works that generally only include a limited number of  lines of interest. The reference articles presented in Tables \ref{table:IDs_HII}, \ref{table:IDs_SFG}, and \ref{table:IDs_PNe} have established the optimal criteria for reddening correction. Most of these studies compare the observed ratio between the fluxes of H$\alpha$ and H$\beta$, or other \hi~lines, with the theoretical expectations, using a calibration curve optimized for the specific conditions of each region. We did not attempt to recalculate the reddening correction in the DESIRED sample, as we consider this parameter to be sufficiently addressed by the cited authors. By default, all spectra considered in DESIRED-E have at least one of the following available temperature diagnostics: \nii~$\lambda 5755/\lambda 6584$, \oiii~$\lambda 4363/\lambda 5007$, or \siii~$\lambda 6312/\lambda 9069$, with uncertainties in the auroral lines of less than 40\%. For this study, we required the direct detection of the \oiii~$\lambda 4363/\lambda 5007$ diagnostic, as well as the detection of \hei~$\lambda 7281$ and \oiii~$\lambda 4363$, with uncertainties of 20\% or less. Additionally, we required the detection of one or both of the following lines: \hei~$\lambda 6678$ and \hei~$\lambda 5876$, which are essential for determining $T_{\rm e}$(\hei). In addition to the aforementioned \hei~RLs, a large portion of the selected observational sample has multiple detections of \hei~singlets with errors of 20\% or less. As part of our study, we also analyzed the fluxes of the \hei~$\lambda 5016$ line in those objects where it is reported.

\begin{figure}
\epsscale{1.15}
\plotone{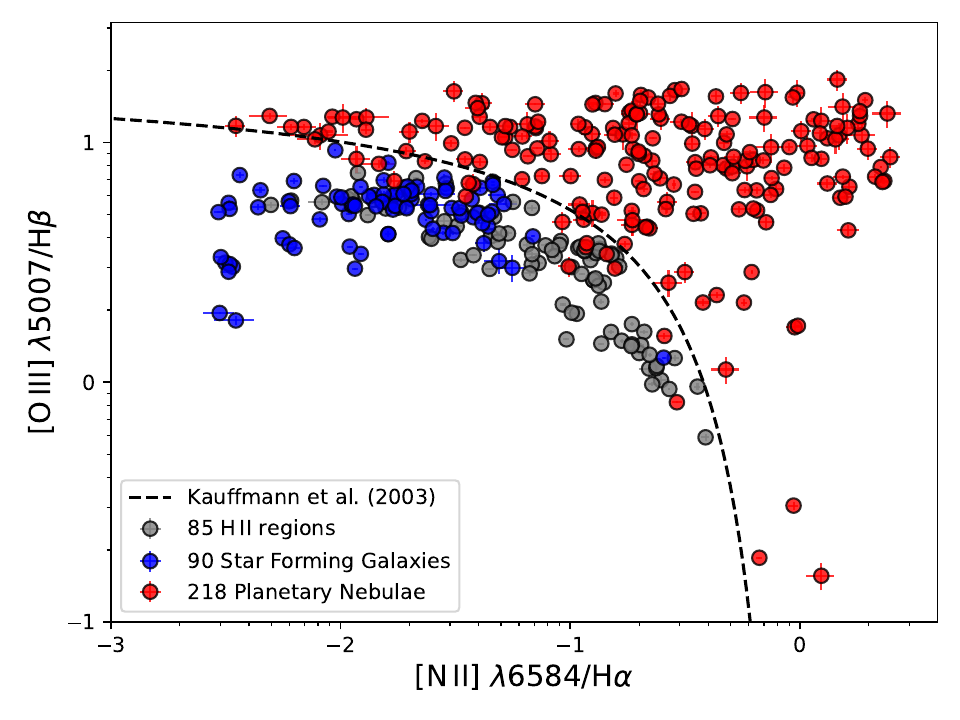}
\caption{BPT diagram of the selected nebular spectra. The dashed line represents the empirical relation by \citet{Kauffmann:03} that distinguishes regions ionized by sources with effective temperatures typical of O and early B type stars (such as star-forming regions) from regions with harder ionizing sources (such as active galactic nuclei or some planetary nebulae).} 
\label{fig:BPT_diagram}
\end{figure}

Our sample of star-forming regions covers the metallicity range from 12+log(O/H)$\approx$7.05 to 12+log(O/H)$\approx$8.60, while the PNe cover from 12+log(O/H)$\approx$7.50 to 12+log(O/H)$\approx$9.10  \citep[derived via the direct method, i.e., $t^2=0$;][]{Peimbert:67}\footnote{The  ``$t^2$'' parameter is the root mean square deviation from the averaged nebular temperature. It is a quantitative measure of the internal temperature variations of the gas, according to the formalism proposed by M. Peimbert. $t^2=0$ implies a homogeneous temperature structure.}. The Baldwin-Phillips-Terlevich (BPT) diagram \citep{Baldwin:81} of the studied sample is shown in Fig.~\ref{fig:BPT_diagram}. 

In this work, we determine $n_{\rm e}$ and $T_{\rm e}$ considering various CEL-diagnostics following the so-called ``direct method'' \citep{Dinerstein:90,Peimbert:17}, with the methodology described in detail in Sec.~\ref{sec:obs} of \citet{MendezDelgado:23b} and \citet{MendezDelgado:24b}. In short, we determine $n_{\rm e}$ considering different diagnostics such as \sii~$\lambda 6731/\lambda6716$, \oii~$\lambda 3726/\lambda3729$, \cliii~$\lambda 5538/\lambda5518$, \feiii~$\lambda 4658/\lambda4702$, and \ariv~$\lambda 4740/\lambda4711$. Then, we adopt an average value considering the following criteria that take into account the different sensitivity ranges of each diagnostic: if $n_{\rm e}$(\sii~$\lambda 6731/\lambda6716) < 100\text{ cm}^{-3}$, we adopt $n_{\rm e}=100 \pm 100\text{ cm}^{-3}$. If $100 \text{ cm}^{-3}\leq n_{\rm e}$(\sii~$\lambda 6731/\lambda6716)$ $< 1000\text{ cm}^{-3}$, we adopt the average between $n_{\rm e}$(\sii~$\lambda 6731/\lambda6716$) and $n_{\rm e}$(\oii~$\lambda 3726/\lambda3729$). If $n_{\rm e}$(\sii~$\lambda 6731/\lambda6716)\geq 1000\text{ cm}^{-3}$, we adopt the averages of $n_{\rm e}$(\sii~$\lambda 6731/\lambda6716$), $n_{\rm e}$(\oii~$\lambda 3726/\lambda3729$), $n_{\rm e}$(\cliii~$\lambda 5538/\lambda5518$), $n_{\rm e}$(\feiii~$\lambda 4658/\lambda4702$), and $n_{\rm e}$(\ariv~$\lambda 4740/\lambda4711$). In some star-forming regions, it was not possible to determine the density; in such cases, we adopted $n_{\rm e}=100 \pm 100\text{ cm}^{-3}$. With the average $n_{\rm e}$, we then calculated $T_{\rm e}$(\oiii~$\lambda 4363/\lambda 5007$). In all cases, we used PyNeb 3.11 \citep{Luridiana:15, Morisset:20}, adopting the atomic data from Table \ref{table:atomic_data}.

The determination of $T_{\rm e}$(\hei) was carried out by comparing the observed intensity ratios of \hei~RLs $\lambda 7281/6678$ and $\lambda 7281/5876$ with theoretical recombination predictions, adopting the average density of each region. As established by \citet{Zhang:05} and \citet{MendezDelgado:21a}, the dependence of these line intensity ratios on $n_{\rm e}$ is small, and potential errors in this parameter introduced by internal density variations \citep{Rubin:89} will not influence our results. Although \hei~$\lambda 5876$ is a line from the triplet system, it is minimally affected by the metastability of the $2^3S$ level. As shown in fig.\ 4 of \citet{Zhang:05} and our Figs \ref{fig:TeHeI_6678_vs_THeI_5876_P1213}, \ref{fig:TeHeI_6678_vs_THeI_5876_DZ22}, there is excellent consistency between $T_{\rm e}$(\hei~$\lambda 7281/\lambda 6678$) and $T_{\rm e}$(\hei~$\lambda 7281/\lambda 5876$). The determination of $T_{\rm e}$(\hei) was made using effective recombination coefficients under the assumption of ``Case B'' from both \citet{Porter:12, Porter:13} and the most recent values from \citet{DelZanna:22} to detect possible errors or anomalies in the atomic data. The first set of atomic data has been widely adopted in the scientific literature for the determination of $Y_p$ \citep{Izotov:14, Valerdi:19, Aver:22} and other studies of He$^{+}$ abundances, and is used by default in the latest version of the Cloudy photoionization model \citep{Ferland:17}. The second set of data has undergone a careful revision of the \hei~model and is expected to be a substantial improvement over previous data.

Unlike the case with CEL-ratios, PyNeb does not have a routine like \textit{getTemDen} to directly determine $T_{\rm e}$ using RL-ratios. For these cases, we define the emissivity of each line using the \textit{getEmissivity} routine from PyNeb over a wide range of temperatures. For the atomic data from \citet{Porter:12, Porter:13}, the temperature range considered is from 5,000K to 25,000K in steps of 10K. The atomic data were not calculated by these authors outside the mentioned temperature range. For the data from \citet{DelZanna:22}, we considered a range from 500K to 32,000K, also using steps of 10K. We then used the \textit{interp1d} function from the SciPy package \citep{Virtanen:20} to interpolate these emissivities and transform the RL-ratios into $T_{\rm e}$ values. In each case, we considered the observational errors of the lines and propagated their effects using Monte Carlo calculations with 1,000 points. The final $T_{\rm e}$(\hei) adopted in this work is the average of $T_{\rm e}$(\hei~$\lambda 7281/\lambda 6678$) and $T_{\rm e}$(\hei~$\lambda 7281/\lambda 5876$), using the atomic data from \citet{DelZanna:22}.

In the case of PNe, we analyze the relationship between the temperature structure, $T_{\rm e}$(\oiii)-$T_{\rm e}$(\hei), and the abundance discrepancy of O$^{2+}$ obtained from RLs and CELs. For the determination of these ionic abundances, $T_{\rm e}$(\oiii) and the averaged density for each region are used in all cases, as well as the reported intensities of \oiii~$\lambda \lambda 4959, 5007$ and the observed RLs from the \oiirls~V1 multiplet ($\lambda \lambda$ 4638.86, 4641.81, 4649.13, 4650.84, 4661.63, 4673.73, 4676.23, 4696.35), following the methodology described in detail by \citet{MendezDelgado:23a}. For this purpose, we adopt the \oiirls~effective recombination coefficients from \citet{Storey:14}.

The densities, temperatures, \hei~line intensities, theoretical \hei~fluxes, and ionic abundances of the analyzed nebulae are presented in Tables from \ref{table:densities_HII} to \ref{table:temps_PNe_toy_model}.

\section{Results}
\label{sec:res}

\begin{figure*}
\epsscale{1.15}
\plotone{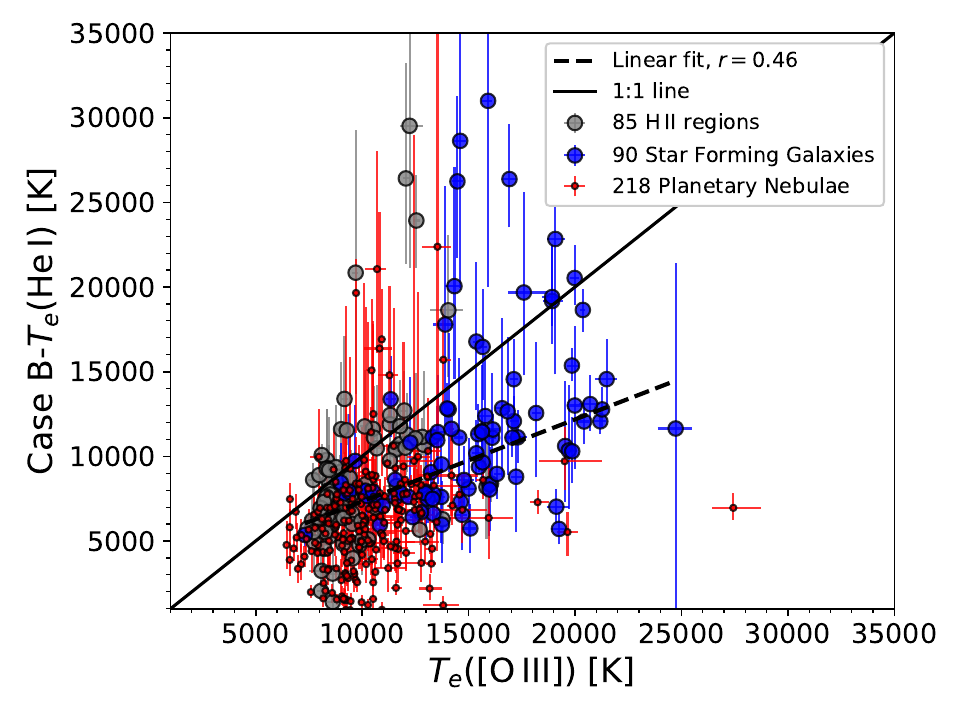}
\caption{Comparison of direct determinations of high-ionization gas temperature diagnostics, $T_{\rm e}$(\hei) and $T_{\rm e}$(\oiii), in several types of photoionized nebulae. $T_{\rm e}$(\hei) was derived by using \hei~$\lambda 7281/ \lambda 5876$ and/or \hei~$\lambda 7281/ \lambda6678$ assuming the ``Case B'' recombination conditions. The linear fit shown considers only \hii~regions and star-forming Galaxies and is presented in Eq.~\eqref{Eq:TeHI_TeO3}.
Note the overall trend of lower values of $T_{\rm e}$(\hei) compared to $T_{\rm e}$(\oiii).}
\label{fig:TeHeI_vs_TOIII}
\end{figure*}

\begin{figure}
\epsscale{1.15}
\plotone{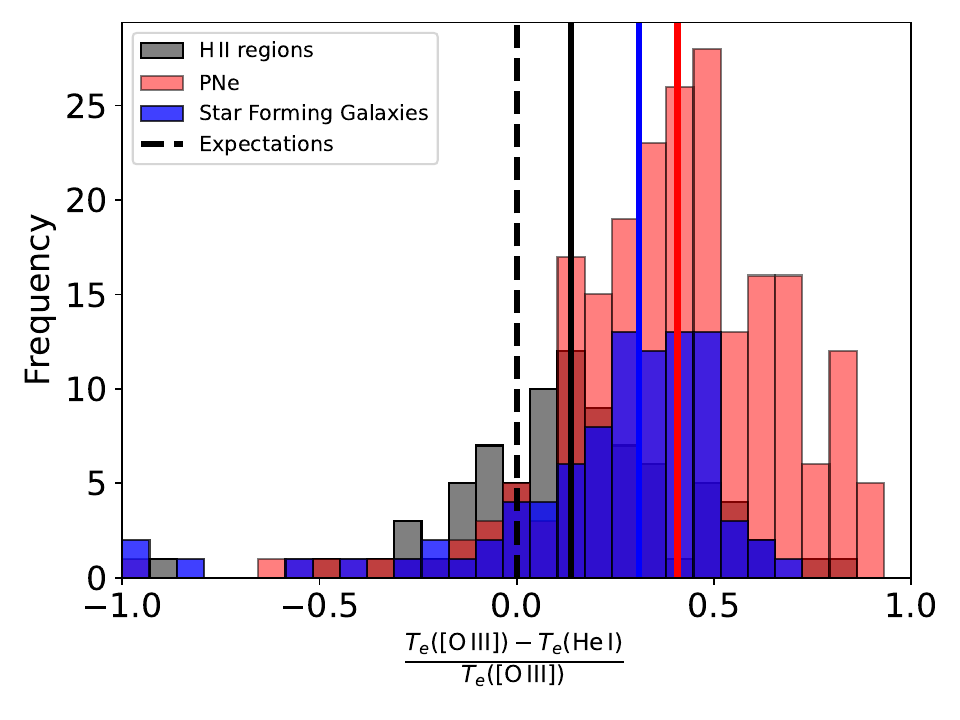}
\caption{Histogram of the differences between $T_{\rm e}$(\oiii) and $T_{\rm e}$(\hei) observed in Fig.~\ref{fig:TeHeI_vs_TOIII}. The colored vertical lines represent the median values of each distribution. The vast majority of all targets show a positive difference with $T_{\rm e}$(\oiii) significantly larger than $T_{\rm e}$(\hei).}
\label{fig:TeHeI_vs_TOIII_hist}
\end{figure}

In Fig.~\ref{fig:TeHeI_vs_TOIII}, we show the resulting distribution of $T_{\rm e}$(\hei) and $T_{\rm e}$(\oiii) for the sample of ionized nebulae analyzed in this work, considering the atomic data from \citet{DelZanna:22}. We have drawn the PNe with smaller symbols than the star-forming regions to avoid overlapping. There is a clear and pronounced trend of $T_{\rm e}$(\hei) $<$ $T_{\rm e}$(\oiii), which cannot be naturally explained by the slight differences in the ionization ranges of He$^+$ and O$^{2+}$, as they predict very small discrepancies of up to $\sim$250~K (see Sec.\ref{sec:appendix_temps}).
Interestingly, qualitatively, the trend observed in star-forming regions appears to be broadly similar to what is obtained in PNe. In the same figure, we have performed a linear fit between $T_{\rm e}$(\hei) and $T_{\rm e}$(\oiii) considering only the star-forming regions, presented in Eq.~\eqref{Eq:TeHI_TeO3}: 
\begin{equation}
\label{Eq:TeHI_TeO3}
T_{\rm e}(\text{\hei})=\left(0.48 \pm 0.05 \right)\times T_{\rm e}(\text{\oiii}) + \left(2500 \pm 600\right) ~[\text{K}].
\end{equation}
The fit was conducted using the Orthogonal Distance Regression (ODR) method, considering uncertainties in both $T_{\rm e}$(\hei) and $T_{\rm e}$(\oiii). The Pearson correlation coefficient $r=0.46$ is moderate due to the observational scatter. However, it is statistically significant ($p < 0.01$) for a sample size of 175 \hii~regions and star-forming galaxies. The histogram of the differences is presented in Fig.~\ref{fig:TeHeI_vs_TOIII_hist}, which shows that the largest deviations occur in planetary nebulae, followed by star-forming galaxies, with \hii~regions being the least discrepant.

We highlight that the observed trend in Fig.~\ref{fig:TeHeI_vs_TOIII} does not depend on the choice of \hei~atomic data. In Fig.~\ref{fig:general_Te_atomic_comparison}, we show that $T_{\rm e}$(\hei) determined using both the atomic data from \citet{Porter:12, Porter:13} and \citet{DelZanna:22} are reasonably consistent\footnote{Note, however, that there is a slight overestimate of $T_{\rm e}$(\hei) when using the data from \citet{Porter:12, Porter:13} in high-temperature regions ($T_{\rm e}>15,000$K). This is due to the overestimate of the emissivity of \hei~$\lambda 6678$ by the atomic data from \citet{Porter:12, Porter:13}, which has been widely discussed and corrected by \citet{DelZanna:22}.}.

\section{Discussion}
\label{sec:disc}

Under ideal conditions of thermal and chemical homogeneity and dominance of the ``Case B'' recombinations, as is generally assumed in the determination of chemical abundances in ionized nebulae, one would expect relatively good consistency between $T_{\rm e}$(\hei) and $T_{\rm e}$(\oiii), given that both are high-ionization ions. In Fig.~\ref{fig:TeHeI_vs_TOIII}, we demonstrate that statistically, there is a general trend of $T_{\rm e}$(\hei)$<$$T_{\rm e}$(\oiii) across all ionized nebulae, including star-forming regions and planetary nebulae, both Galactic and extragalactic. This result is very important because it indicates a serious issue in one of the basic physical principles assumed to infer $T_{\rm e}$(\hei) and/or $T_{\rm e}$(\oiii).

\subsection{Are There Potential Systematic Errors in $T_{\rm e}$(\hei)?
}
\label{sec:TeHeI_systematics}

\begin{figure}
\epsscale{1.15}
\plotone{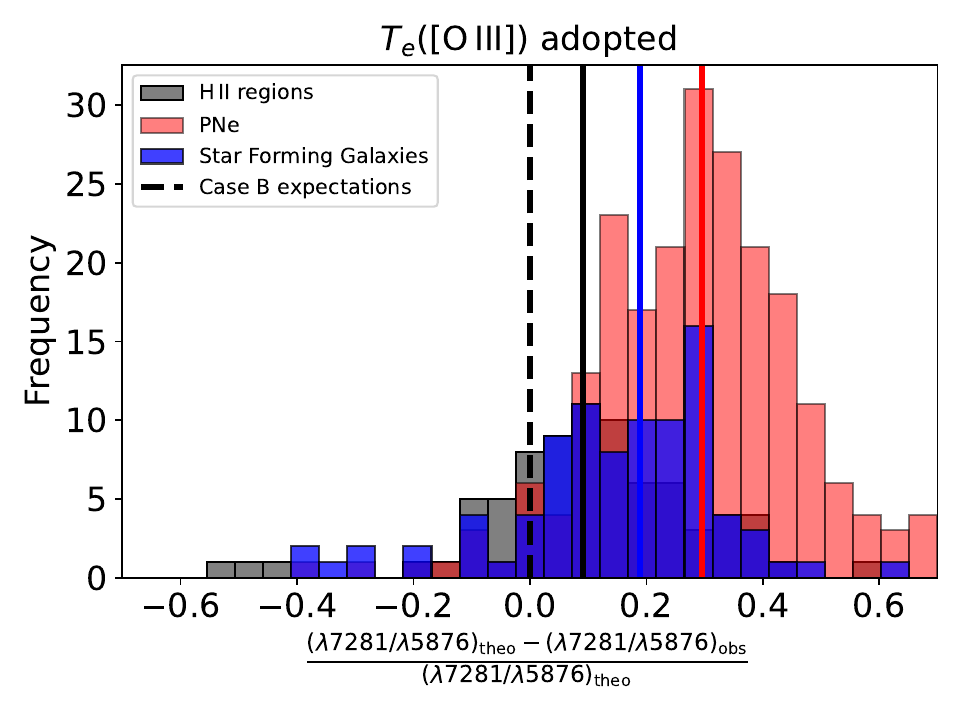}
\caption{Distribution of the differences between the theoretical and observed values of the intensities of the \hei~$\lambda 7281$ line, using the flux of the \hei~$\lambda 5876$ line as normalization in the sample of analyzed regions. $T_{\rm e}$(\oiii) was adopted to model the line emissivities under ``Case B'' using the recombination coefficients from \citet{DelZanna:22}.The colored vertical lines represent the median values of each distribution.}
\label{Fig:distribution_7281_TO3}
\end{figure}

$T_{\rm e}$(\hei) was determined from the line ratios of \hei~$\lambda 7281/\lambda 6678$ and \hei~$\lambda 7281/\lambda 5876$. If $T_{\rm e}$(\hei) is ``too low'' it could be because the flux of \hei~$\lambda 7281$ is fainter or the flux of \hei~$\lambda \lambda 6678, 5876$ is brighter than expected. Quantitatively, in star-forming regions, the flux of \hei~$\lambda 7281$ should be systematically about $\sim$15\% more intense than what we observe to achieve good consistency between $T_{\rm e}$(\hei) and $T_{\rm e}$(\oiii), as shown in Fig.~\ref{Fig:distribution_7281_TO3}. For PNe, the difference is approximately $\sim$30\%. One might wonder if this is simply an error in the effective recombination coefficients for this line. This hypothesis seems unlikely considering that almost all available atomic calculations in the literature, using various approaches and independently, converge on very similar emissivities for this line \citep{Smits:96, Benjamin:99, Porter:05, Porter:12, Porter:13, DelZanna:22}.

Alternatively, there could be errors in the emissivities of \hei~$\lambda 6678$ and \hei~$\lambda 5876$. This hypothesis would call into question most determinations of He$^{+}$ abundance in the literature, which are generally based on the intensity of the brighter optical \hei~lines, particularly \hei~$\lambda 5876$. Nevertheless, the good consistency shown between observational and theoretical values of the relative fluxes of \hei~$\lambda 5876/ \lambda 6678$ (see Fig.~\ref{fig:TeHeI_6678_vs_THeI_5876_DZ22}), despite coming from systems with different spin multiplicity, seems to rule out this idea.

The fact that the observed flux of \hei~$\lambda 7281$ is significantly fainter than expected from recombination theory was noted by \citet{Porter:07} in the observations of the Orion Nebula reported by \citet{Esteban:04}. In the face of the mismatch between the model and observations for this line, those authors suggested that the error must come from the observations, particularly from the reddening correction adopted by \citet{Esteban:04}, which was based on the work of \citet{Costero:70}. Certainly, extinction plays an important role when comparing the relative fluxes to H$\beta$ of the different \hei~lines, and in this sense, the use of the extinction curve by \citet{Blagrave:07} in the Orion Nebula induces a notable improvement\footnote{In this work, we use the fluxes of the Orion Nebula from \citet{Esteban:04}, corrected for reddening using the curve from \citet{Blagrave:07} instead of that from \citet{Costero:70} and the \hi~Balmer and Paschen lines \citep{MendezDelgado:23b}. As shown in Table~\ref{table:temps_HII}, the ``problem'' with \hei~$\lambda 7281$ is not eliminated; it is only reduced by 6\% with respect to the use of the reddening curve from \citet{Costero:70}. }. However, the role that optical extinction plays in the \hei~$\lambda 7281/\lambda 6678$ and \hei~$\lambda 7281/\lambda 5876$ ratios is very small. More importantly, to explain the general trend observed in Fig.~\ref{fig:TeHeI_vs_TOIII} in terms of errors in the extinction correction, there would need to be a widespread error in all the reddening curves used in the spectral range around $\sim \lambda 7281$, regardless of the geometric, dust, and metallicity conditions of the ionized regions studied here, which is highly unlikely.

Other errors in \hei~$\lambda 7281$, such as contamination by sky lines, which are abundant in this spectral region, may still be present, even though the DESIRED project carefully noted observational defects to avoid introducing spurious noise. This contamination could artificially increase the measured flux of \hei~$\lambda 7281$, thus raising $T_{\rm e}$(\hei), which might explain why a small subset of regions show $T_{\rm e}$(\hei) $>$ $T_{\rm e}$(\oiii) in Fig.~\ref{fig:TeHeI_vs_TOIII} despite visual checks. Hypothetically speaking, the presence of telluric absorptions in \hei~$\lambda 7281$ could induce the observed effect of $T_{\rm e}$(\hei) $<$ $T_{\rm e}$(\oiii). However, our sample includes both Galactic and extragalactic objects with different radial velocities and spectral resolutions, making it unlikely that a systematic effect, such as sky contamination or telluric absorptions, would affect all objects in the same way. This effect alone cannot explain the general trend that, statistically, $T_{\rm e}$(\hei) $<$ $T_{\rm e}$(\oiii).

\begin{figure}
\epsscale{1.15}
\plotone{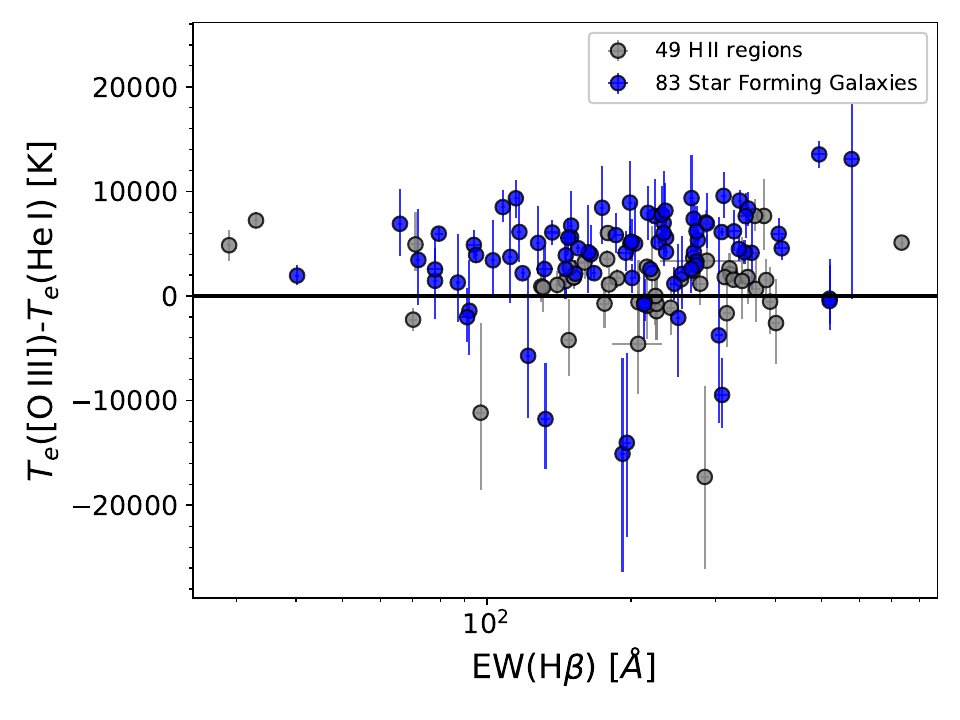}
\caption{Comparison between the temperature difference $T_{\rm e}$(\oiii) - $T_{\rm e}$(\hei) and the reported equivalent width of H$\beta$ (EW(H$\beta$)). If the temperature differences were due to uncorrected stellar absorption in the \hei~lines, a trend with EW(H$\beta$) would be expected. The effects of stellar absorption should be more pronounced in regions with lower EW(H$\beta$).}
\label{fig:TeEW}
\end{figure}

Underlying stellar absorption can significantly affect \hei~emission lines, reducing their observed flux, as demonstrated in studies on primordial helium abundance \citep{Skillman:1998, Olive:2001, Peimbert:07, Aver:15}. If absorption is more significant in \hei~$\lambda 7281$ than in \hei~$\lambda 5876$ or \hei~$\lambda 6678$, it could artificially lower $T_{\rm e}$(\hei). To test this in our extragalactic \hii~regions and star-forming galaxies, we compared the $T_{\rm e}$(\oiii)$-$$T_{\rm e}$(\hei) difference with the equivalent width of H$\beta$ (EW(H$\beta$)), as stellar absorption is expected to be more important in regions with low EW(H$\beta$) \citep{Izotov:04}. Fig.~\ref{fig:TeEW} shows no clear correlation between $T_{\rm e}$(\oiii)$-$$T_{\rm e}$(\hei) and EW(H$\beta$), suggesting that stellar absorption does not cause the trend in Fig.~\ref{fig:TeHeI_vs_TOIII}. Additionally, Galactic \hii~regions and PNe, where stellar absorption is absent, also show the same pattern of $T_{\rm e}$(\hei)$<$$T_{\rm e}$(\oiii) found in the extragalactic star-forming regions.

Finally, it is important to note that a hypothetical metastability of the $2^1S$ level, analogous to the $2^3S$ level in the triplet system, would cause lines such as \hei~$\lambda 5016$ to be less intense than their theoretical recombination predictions, while \hei~$\lambda 7281$ would show the opposite effect, being more intense than predicted. This scenario is not observed and is therefore ruled out.

\subsection{Assessing the Validity of the ``Case B'' Assumption for the Singlet \hei~Atom}
\label{sec:CaseB}

The fact that \hei~$\lambda 7281$ is fainter than predicted by recombination theory could indicate that some process is reducing the population of the $3^1S$ level. In addition to direct recombination to this level, it is substantially populated through transitions from the $n^1P$ levels, as shown in Fig.~\ref{fig:grotrian}. 

Among all $n^1P\rightarrow3^1S$  transitions, the most important are those from the $4^1P$ and $5^1P$ levels, which give rise to the infrared lines \hei~$\lambda 15084$\AA~and $\lambda 11016$\AA, respectively \footnote{Note that these lines have been scarcely studied in the literature, but they fall within the spectral range covered by instruments like NIRSPEC on the James Webb Space Telescope (JWST). In the presence of temperature variations and/or deviations from Case B recombination, these lines should also be affected in a manner analogous to the rest of the \hei~singlets studied in the present article.}. These transitions contribute approximately 30\% and 20\% of the flux from all lines emitted by transitions from various P levels to the $3^1S$ level \citep{DelZanna:22}. If the rate of $n^1P \rightarrow 3^1S$ transitions decreases for some reason, the population of the $3^1S$ level will also decrease significantly, and consequently, the flux of \hei~$\lambda 7281$ will be lower than expected.

\begin{figure}
\epsscale{1.15}
\plotone{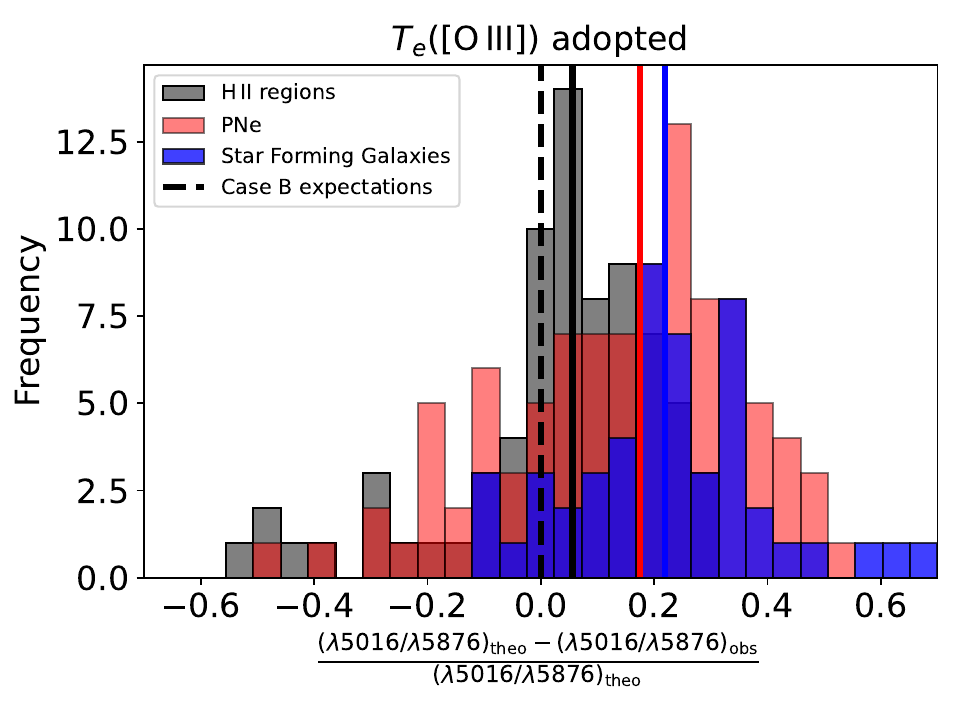}
\caption{The distribution of the differences between the theoretical and observed values of the intensities of the \hei~$\lambda 5016$ line, using the flux of the \hei~$\lambda 5876$ line as normalization in the sample of analyzed regions. This is the same plot as in Fig.~\ref{Fig:distribution_7281_TO3} but considering the \hei~$\lambda 5016$ line instead of the \hei~$\lambda 7281$ line.}
\label{Fig:distribution_5016_TO3}
\end{figure}

\begin{figure}[ht!]
\epsscale{1.15}
\plotone{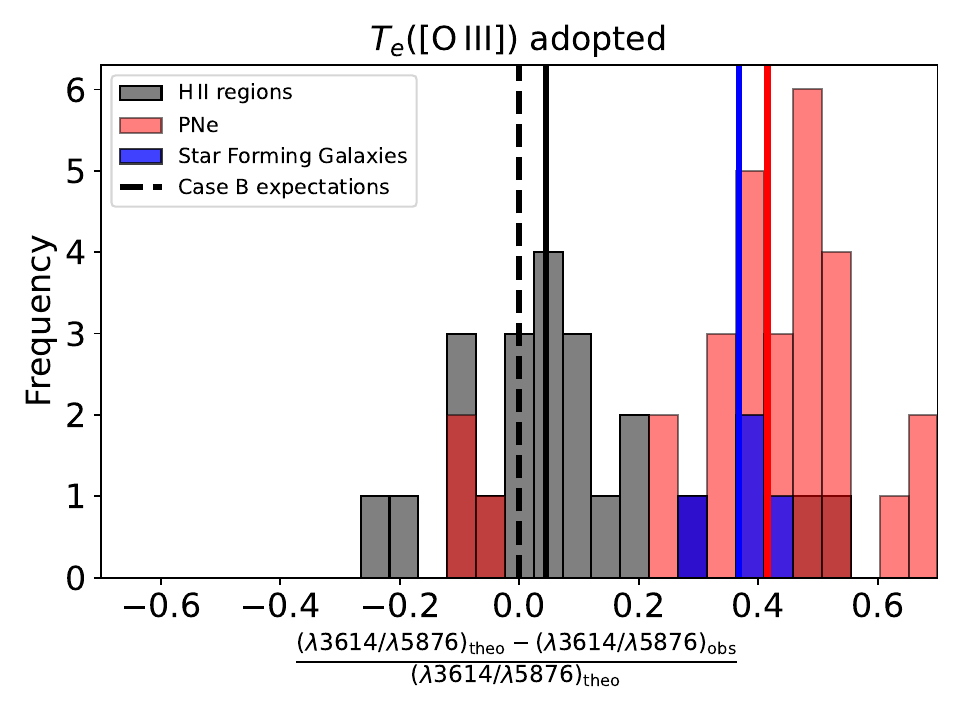}
\caption{Same as in Fig.~\ref{Fig:distribution_7281_TO3} but considering the \hei~$\lambda 3614$ line.}
\label{Fig:distribution_3614_TO3}
\end{figure}

\begin{figure}[ht!]
\epsscale{1.15}
\plotone{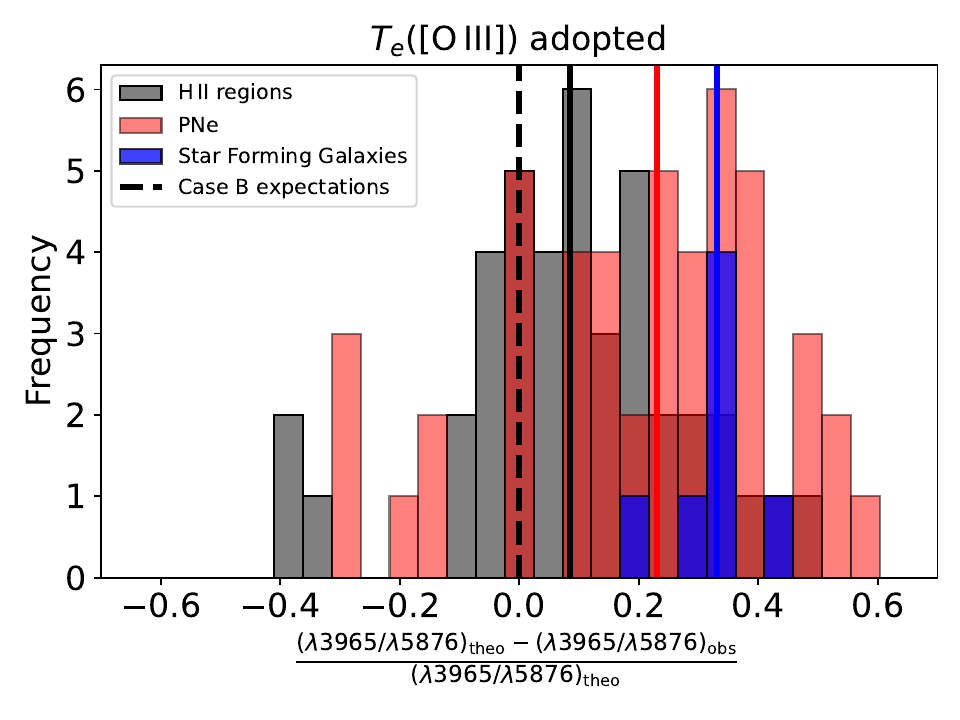}
\caption{Same as in Fig.~\ref{Fig:distribution_7281_TO3} but considering the \hei~$\lambda 3965$ line.}
\label{Fig:distribution_3965_TO3}
\end{figure}

A phenomenon that could cause a decrease in the rate of $n^1P \rightarrow 3^1S$ transitions is a general reduction in the electron population of the $n^1P$ levels. This could occur if, on average, photons originating from the transitions $n^1P \rightarrow 1^1S$, such as those marked in red in Fig.~\ref{fig:grotrian}, ``escape'' or are absorbed by other ions or compounds rather than being reabsorbed by \hei. This would violate the ``Case B'' recombination assumption, and in the extreme case where all photons from $n^1P \rightarrow 1^1S$ are lost, the conditions would resemble those of ``Case A''\footnote{Note that the fact that photons from $n^1P \rightarrow 1^1S$ could be escaping on average does not imply the absence of potential fluorescent excitations $1^1S \rightarrow n^1P$ coexisting with the aforementioned photon escaping. This is known as ``Case C'' \citep{Baker:38,Ferland:99}.}. Obviously, ``Case A'' generally does not apply for \hei~because it would imply that the emissivity of the \hei~singlets arising from the $n^1P$ levels should be up to two orders of magnitude smaller than in ``Case B''\citep{Brocklehurst:72,Smits:96}, and therefore lines such as \hei~$\lambda \lambda 5016, 3614, 3965$ should not be widely detected at the observed fluxes. In reality, these lines are observed with fluxes lower than expected but not as much as in the ``Case A'', as shown in Figs \ref{Fig:distribution_5016_TO3}, \ref{Fig:distribution_3614_TO3}, and \ref{Fig:distribution_3965_TO3}. An interesting hypothetical case to explore would be an intermediate situation in which a fraction of the singlet \hei~is in conditions close to ``Case A'' while the rest is in ``Case B''.

\subsection{A toy model with a mixed Case A+Case B scenario}
\label{sec:toy_model}

Now, let us explore the case where the differences between the observed fluxes and the theoretical predictions presented in Figs. \ref{Fig:distribution_7281_TO3}, \ref{Fig:distribution_5016_TO3}, \ref{Fig:distribution_3614_TO3}, and \ref{Fig:distribution_3965_TO3} are assumed to be due to deviations from ``Case B'' as some authors have suggested \citep{Liu:06,Izotov:07}. For this purpose, let us assume the following toy model:

\begin{equation}
    \label{eq:caseB_caseA_fraction}
    I(\lambda)_{\text{obs}} = \gamma \times I(\lambda)_{\text{Case B}} + (1 - \gamma) \times I(\lambda)_{\text{Case A}},
\end{equation}

where $I(\lambda)_{\text{obs}}$ represents the total observed intensity of a \hei~singlet line, $I(\lambda)_{\text{Case B}}$ is the intensity that would be emitted under ``Case B'' and $\gamma$ is the fraction of the gas that is in ``Case B'' conditions. To determine $\gamma$ in each case is relatively straightforward.  One can use the fact that the ``Case B'' and ``Case A'' emissivities of \hei~$\lambda 5016$ differ substantially and to a greater extent than \hei~$\lambda 7281$, as do the other transitions originating from the $n^1P$ levels. We can use the intensity of the \hei~$\lambda 5876$ line to normalize the emissivities, as it is a triplet case-independent line with very small effects from self-absorption. Considering Eq.~\ref{eq:caseB_caseA_fraction} for the specific case of \hei~$\lambda 5016$, we derive Eq.~\ref{eq:caseB_caseA_fraction_5016}:

\begin{equation}
    \label{eq:caseB_caseA_fraction_5016}
    \frac{I(5016)_{\text{obs}}}{I(5876)_{\text{obs}}}\times\frac{j(5876)_{\text{Case B}}}{j(5016)_{\text{Case B}}}=\gamma + \epsilon \times \left(1-\gamma\right),
\end{equation}

where $j(\lambda)$ represents the emissivity of the line at $\lambda$, and $\epsilon = j(5016)_{\text{Case A}} / j(5016)_{\text{Case B}}$. The precise value of $\epsilon$ depends very little on the physical conditions of the gas and is approximately $\epsilon \approx 2.30 \times 10^{-2}$ \citep{Smits:96}. 

Using the observational values of \hei~$\lambda 5016$ from our sample of ionized nebulae and adopting $T_{\rm e}$(\oiii) by default, we determine $\gamma$ following Eq.~\ref{eq:caseB_caseA_fraction_5016}. Subsequently, we use this value of $\gamma$ to model the emissivities of \hei~$\lambda 7281$ and \hei~$\lambda 6678$, considering Eq.~\ref{eq:caseB_caseA_fraction}, although for the latter line, the difference between ``Case A'' and ``Case B'' is around 1\%. Note that for this, we need to use the ``Case A'' recombination coefficients from \citet{Smits:96}, as the calculations by \citet{Porter:12,Porter:13} and \citet{DelZanna:22} are limited to ``Case B''. This may introduce some additional uncertainties. To mitigate them, instead of directly combining the predictions of $j(\lambda)_{\text{Case A}}$ from \citet{Smits:96} along with the predictions of $j(\lambda)_{\text{Case B}}$ from \citet{DelZanna:22}, we use $j(\lambda)_{\text{Case B, DZ22}} \times j(\lambda)_{\text{Case A, S96}} / j(\lambda)_{\text{Case B, S96}}$. Although the emissivities for \hei~$\lambda 5016$, \hei~$\lambda 7281$, and \hei~$\lambda 6678$ may differ slightly between \citet{Smits:96} and \citet{DelZanna:22}, the relative ratios of the emissivities between the two recombination cases are in excellent agreement.  This can be seen in Table~A1 of \citet{DelZanna:22}, which shows the comparison of the emissivities between ``Case A'' and ``Case B'' for the strongest \hei~lines at $T_{\rm e} = 20,000 \text{ K}$ and $n_{\rm e} = 10^6 \text{ cm}^{-3}$.

\begin{figure}[h]
\epsscale{1.15}
\plotone{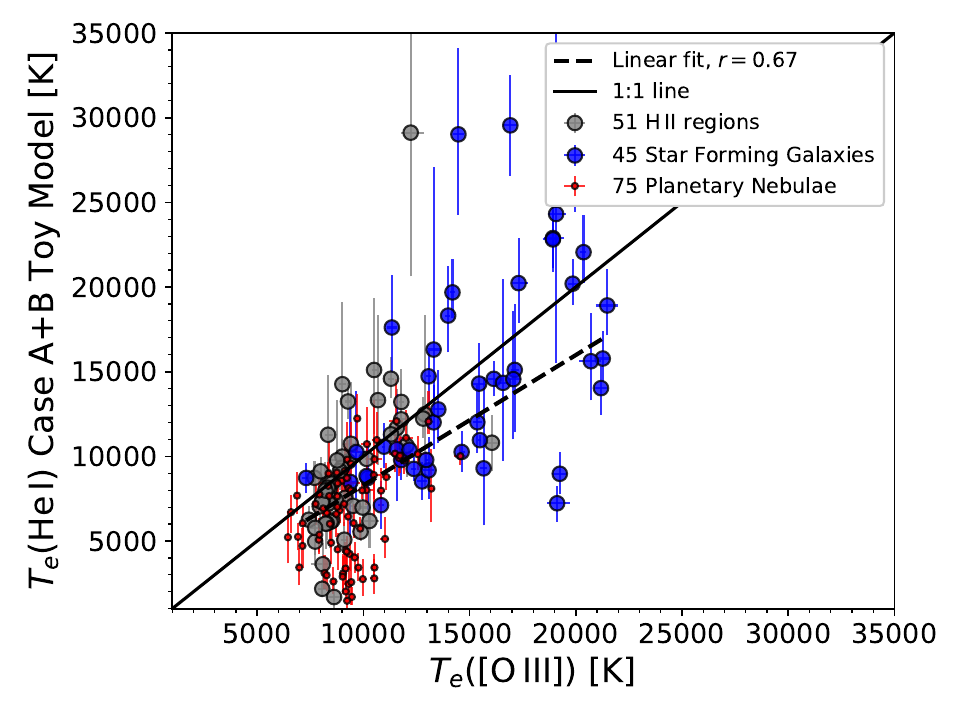}
\caption{The same as Fig.~\ref{fig:TeHeI_vs_TOIII}, but considering a combination of ``Case B'' and ``Case A'' conditions in a toy model described by Eq.~\eqref{eq:caseB_caseA_fraction}. The linear fit shown considers only \hii~regions and star-forming Galaxies and is presented in Eq.~\eqref{Eq:eqtoymodel}. }
\label{fig:ToyModel}
\end{figure}

The median $\gamma$ values obtained for each type of object are 0.75, 0.88, and 0.75 for PNe, HII regions, and SFGs, respectively. Using the toy model described above, we can re-determine $T_{\rm e}$(\hei\ $\lambda 7281 / \lambda 6678$) and $T_{\rm e}$(\hei\ $\lambda 7281 / \lambda 5876$) using the same procedure described in Sec~\ref{sec:obs}, now adding the subscript $TM$. In Fig.~\ref{fig:ToyModel}, we show the resulting distribution from this toy model, where $T_{\rm e}(\text{\hei})_{TM} < T_{\rm e}(\text{\oiii})$, although the difference is substantially smaller than that shown in Fig.~\ref{Eq:TeHI_TeO3}. The linear fit considering only the star-forming regions is shown in Eq.~\eqref{Eq:eqtoymodel}:
\begin{equation}
\label{Eq:eqtoymodel}
T_{\rm e}(\text{\hei})_{TM}=\left(0.77 \pm 0.08 \right)\times T_{\rm e}(\text{\oiii}) + \left(540 \pm 940\right) ~[\text{K}].
\end{equation}
The slope in Eq.~\eqref{Eq:eqtoymodel} is closer to one than in Eq.~\eqref{Eq:TeHI_TeO3} and the offset is significantly smaller.  However, a discrepancy still exists.

\begin{figure}
\epsscale{1.15}
\plotone{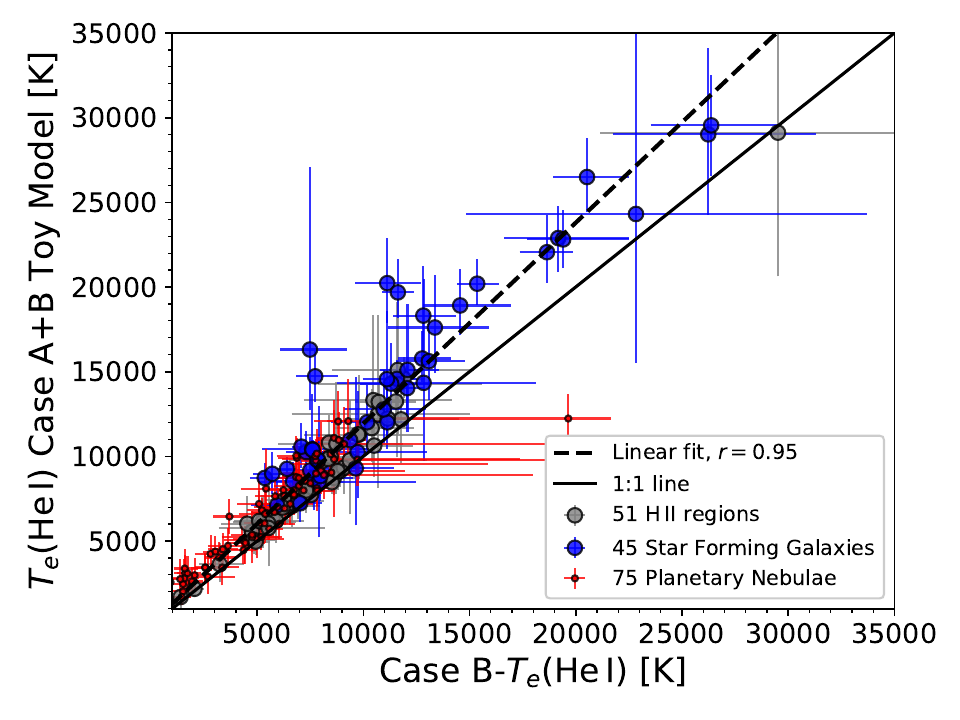}
\caption{Comparison between $T_{\rm e}$(\hei) obtained assuming the ``Case B'' recombination versus $T_{\rm e}$(\hei)$_{TM}$, calculated assuming a mixture between ``Case A'' and ``Case B'', as described in Sec.~\ref{sec:toy_model}. The linear fit considering all spectra is shown in Eq.~\eqref{Eq:eqtoymodel_THeI_vs_CaseB}.}
\label{Fig:toy_model_TheI_vs_caseB}
\end{figure}

The comparison between $T_{\rm e}(\text{\hei})$ and $T_{\rm e}(\text{\hei})_{TM}$ is shown in Fig.~\ref{Fig:toy_model_TheI_vs_caseB}, and the fitted relationship is presented in Eq.~\eqref{Eq:eqtoymodel_THeI_vs_CaseB}:
\begin{equation}
\label{Eq:eqtoymodel_THeI_vs_CaseB}
T_{\rm e}(\text{\hei})_{TM}=(1.19 \pm 0.03 )\times T_{\rm e}(\text{\hei})_{\text{Case B}} + (70 \pm 170) ~[\text{K}].
\end{equation}
Here, the slope is actually greater than one but the offset has dimished considerably.

The fact that $T_{\rm e}(\text{\hei})_{TM} < T_{\rm e}(\text{\oiii})$ implies a failure of the initial hypothesis that the differences between the theoretical and observed values of \hei~singlets were, in general, due to deviations from ``Case B'' to ``Case A''. This demonstrates that this toy model cannot completely explain Fig.~\ref{fig:TeHeI_vs_TOIII}. Even if there are general deviations from ``Case B'' to ``Case A'', one would need to invoke temperature variations or another additional phenomenon to achieve simultaneous consistency between the observed and predicted intensities of \hei~$\lambda 5016$ and \hei~$\lambda 7281$ in all the analyzed objects. Possibly, a complete radiative treatment, such as in ``Case C'' where there is a fluorescent excitation component, could be useful for exploring this. In principle, the photoionization code Cloudy \citep{Ferland:17} includes a complete radiative treatment and its models include predictions ``with all processes included''. However, as demonstrated by \citet{Izotov:13}, this treatment appears inconsistent with observational values for \hei~ and remains as an open issue. A comprehensive radiative treatment requires studying each object individually, which is well beyond the scope of this work.

\begin{figure}
\epsscale{1.15}
\plotone{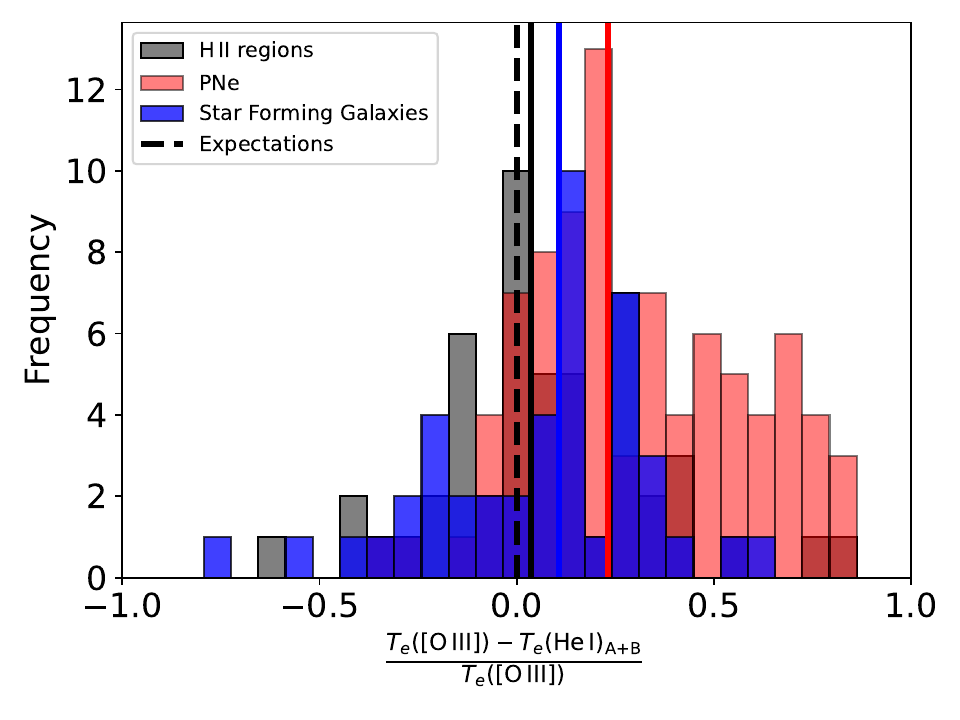}
\caption{Same as Fig.~\ref{fig:TeHeI_vs_TOIII_hist} but considering $T_{\rm e}$(\hei)$_{\text{A+B}}$, determined using the Toy Model of Eq.~\eqref{eq:caseB_caseA_fraction}.}
\label{fig:TeHeI_vs_TOIII_hist_AB}
\end{figure}

If we perform a separate analysis for the different objects, creating a histogram similar to the one presented in Fig.~\ref{fig:TeHeI_vs_TOIII_hist} but using $T_{\rm e}$(\hei)$_{\text{A+B}}$ determined from the combination of Cases A + B in this toy model, we obtain the distribution shown in Fig.~\ref{fig:TeHeI_vs_TOIII_hist_AB}. In this figure, we observe that in \hii~regions, the differences between $T_{\rm e}(\text{\hei})_{\text{A+B}}$ and $T_{\rm e}(\text{\oiii})$ are small, around $\sim$4\%, indicating good consistency. However, this is not the case for the other objects: SFG show a difference of  $\sim$11\%, while PNe exhibit a difference of $\sim$23\%. This could suggest that deviations from ``Case B'' to ``Case A'' alone might provide a reasonable explanation for the observed discrepancy between $T_{\rm e}(\text{\hei})_{\text{B}}$ and $T_{\rm e}(\text{\oiii})$ in \hii~regions, but not for the other objects. However, \hii~regions only cover a narrow range of $T_{\rm e}(\text{\oiii})$ temperatures. Therefore, it can not be discarded that these regions achieve consistency simply because they lie within the natural intersection zone of Eq.~\eqref{Eq:eqtoymodel} with the 1:1 line.

Given the previous discussion, we caution against using \hei~singlet lines for extinction corrections as is suggested by some authors \citep{Zamora:22}. Relying on their observed fluxes for such corrections while assuming a ``Case B'' scenario will lead to systematic errors.

\subsection{Are the temperature variations sufficient to simultaneously explain \hei~$\lambda 5016$ and \hei~$\lambda 7281$?}
\label{sec:t2_self_test}

In Sec.~\ref{sec:toy_model} we have demonstrated that ``Case B'' deviations alone are not sufficient to simultaneously explain, in a quantitatively way, the low observed fluxes of \hei~$\lambda 5016$ and \hei~$\lambda 7281$ compared to what is expected when adopting $T_{\rm e}$(\oiii)--perhaps, the exception of the \hii~regions. Now it is time to explore if $T_{\rm e}$(\hei) derived assuming ``Case B'' and the ratios $\lambda 7281/\lambda 6678$ and/or $\lambda 7281/\lambda 5876$ could be representative of the He$^+$ ion. Under this premise, one would expect the theoretical predictions of \hei~$\lambda 5016/\lambda 5876$, assuming $T_{\rm e}$(\hei), to be consistent with the observed values. If this premise holds true, it could be interpreted as suggesting that the temperature variations, caused by some physical phenomenon, are sufficient to explain the trend of $T_{\rm e}$(\hei) $<$ $T_{\rm e}$(\oiii) observed in Fig.~\ref{fig:TeHeI_vs_TOIII}.

\begin{figure}
\epsscale{1.15}
\plotone{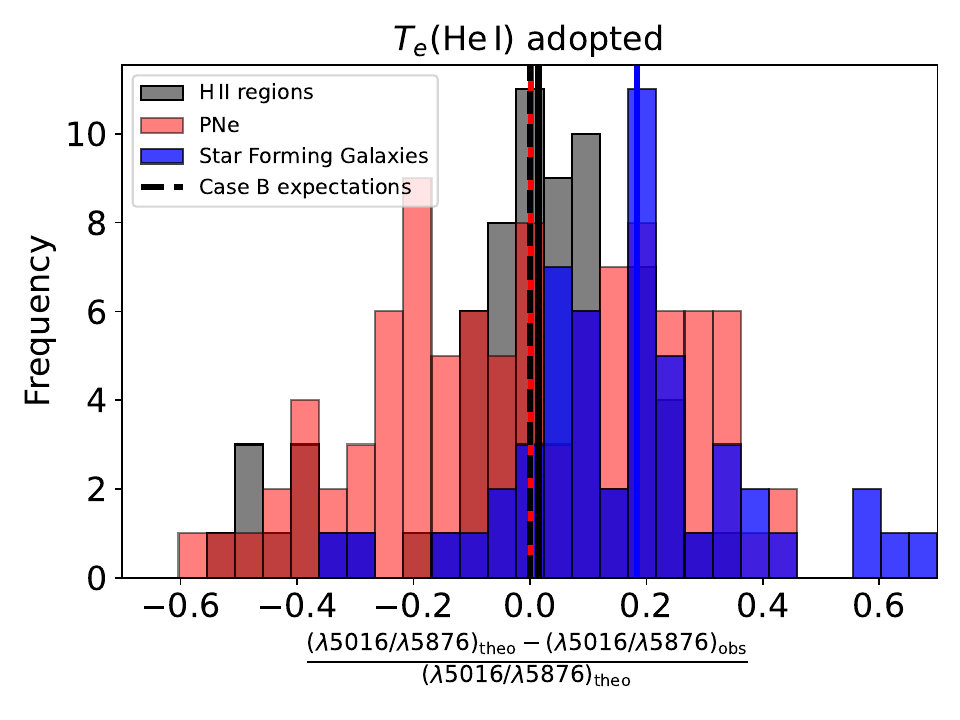}
\caption{Same as in Fig.~\ref{Fig:distribution_7281_TO3}, but considering the \hei~$\lambda 5016$ line and $T_{\rm e}$(\hei). This temperature was derived from the \hei~$\lambda 7281/\lambda 5876$ and/or $\lambda 7281/\lambda 6678$ ratios under ``Case B", using the recombination coefficients from \citet{DelZanna:22}.}
\label{Fig:distribution_5016_THeI}
\end{figure}

In Fig.~\ref{Fig:distribution_5016_THeI}, we show that, statistically, there is good consistency between the predictions and observations of \hei~$\lambda 5016/5876$ in \hii~regions and PNe, but not in SFGs, when adopting $T_{\rm e}$(\hei) derived under ``Case B'' conditions. This suggests that, in the first two groups of objects, $T_{\rm e}$(\hei) might be representative of the He$^+$ ion, and its discrepancy with $T_{\rm e}$(\oiii) could reveal the presence of temperature inhomogeneities caused by gas heating or the presence of cold, high-metallicity clumps, with only a minor impact from potential deviations from ``Case B''. However, the broad width in the distribution of observed values does not allow us to rule out the possibility that deviations from ``Case B'' may still be generally relevant. In the case of SFGs, even when considering $T_{\rm e}$(\hei), the observed values of $\lambda 5016/5876$ are statistically too low. This indicates that, although temperature variations may be present, an additional phenomenon is required to achieve simultaneous consistency between the predictions and observations of the $\lambda 7281$ and $\lambda 5016$ singlets.

\subsection{Potential Causes and Consequences of Partial Deviations from ``Case B''}
\label{sec:CaseB_consequences}

As discussed in Sec.~\ref{sec:toy_model}, the deviations from the ``Case B'' recombination towards the ``Case A'' as modeled in Eq.~\ref{eq:caseB_caseA_fraction} does not seem to explain, on its own, the observed trend of $T_{\rm e}$(\hei) $<$ $T_{\rm e}$(\oiii) for most objects. However, it is likely to contribute substantially to the reduction in the flux of \hei~$\lambda 7281$. This phenomenon appears to be generally present in all the ionized regions analyzed here, including Galactic and extragalactic \hii~regions, star-forming galaxies, and planetary nebulae. In the sample analyzed, factors such as geometry, dust abundance, and velocity gradients —physical conditions that might induce deviations from the ``Case B'' recombination scenario towards ``Case A'' \citep{Cota:88}— are expected to vary widely. It is difficult to identify a single physical mechanism that could account for this recombination case deviation, given that it must be present across all ionized nebulae.

Helium accounts for approximately $\sim$10\% of the total number of atoms in the gas of ionized nebulae. If around $\sim$25\% of \hei~atoms in the singlet system are losing $n^1P \rightarrow 1^1S$ photons, this number of atoms is comparable to the total number of all metals combined. This makes it unlikely that an specific heavy element ion could completely absorb these photons before they are reabsorbed by helium, although certainly many heavy elements could be excited or ionized by these photons, as they have energies between 21.2 and 24.6 eV. Given that dust abundance within photoionized environments changes with metallicity, dramatically decreasing in metal-poor regions \citep{Roman-Duval:22, Roman-Duval:22a, MendezDelgado:24b}, it is unlikely that dust alone accounts for all the missing photons. This is especially true in metal poor star-forming galaxies, which have lower dust-to-gas ratios, and yet exhibit significant discrepancies in \hei~$\lambda \lambda 5016, 7281$ (see Figs.~\ref{Fig:distribution_5016_TO3} and \ref{Fig:distribution_7281_TO3}). The only element that appears sufficiently abundant to absorb such photons is \hi.

The photons from the $n^1P \rightarrow 1^1S$ transitions of the singlet system of \hei~are able to photoionize \hi. This suggests that these photons could be lost by being absorbed through the photoionization process of \hi, rather than being reabsorbed by \hei. The photoionization cross-section of \hi~is quite high around $\sim$13.6 eV but decreases rapidly at higher energies \citep{Bell:67, Brown:71}. The probability that \hi~is ionized by a photon of $\sim$22 eV is approximately $\sim4$ times less than being ionized by a photon near 13.6 eV, which makes this scenario seem plausible as the \hi~is around $\sim9$ times more abundant than \hei. If this mechanism is indeed operating, it would imply the need to include complete radiative treatments that consider these photoionizations of \hi~originating from \hei, which current photoionization models do not naturally predict. Additionally, the possibility that $n^1P \rightarrow 1^1S$ photons could partially excite permitted transitions in a wide variety of ions should not be dismissed \citep{ReyesRodriguez:24} and remains an open question for several elements.

Another possibility, though more speculative, is the escape of these ionizing photons into the ISM without being degraded into lower-energy photons. Since helium is the second most abundant element in the universe, the quantity of ionizing photons contributing to the ISM would be significant enough to play an important role in the presence of the Diffuse Ionized Gas (DIG). The DIG is observed in nearly all galaxies, and its origin is widely debated \citep{Wood:04, Haffner:09, Belfiore:22, McClymont:24, GonzalezDiaz:24a, GonzalezDiaz:24b}. The escape of relatively energetic photons, such as the \hei~$n^1P \rightarrow 1^1S$ photons, could contribute to the relatively hard ionization conditions of the DIG \citep{Wood:04}.

In the speculative case of a \hei~$n^1P \rightarrow 1^1S$ photon escaping from the ionized ISM, one might wonder if such escape could also occur for \hi, an atom with similar configuration.  The emissivity of some \hi~RLs, such as H$\alpha$ or H$\beta$, under ``Case A'' recombination is approximately 65\% lower than the predictions under ``Case B'' \citep{Storey:95}. If there is a $\sim$25\% deviation from ``Case B'' to ``Case A'' for \hi, the overall emissivity of these \hi\ RLs would decrease by roughly $\sim$10\%. This would imply a general systematic error in the chemical abundances derived from CELs or RLs that are independent of the recombination case with respect to \hi, resulting in overestimations by the same fraction of about $\sim$10\%. For example, the determinations of He/H abundances obtained from \hei~RLs of the triplet system, as is generally the case in the literature \citep{Izotov:07, Aver:15, Valerdi:19}, would be systematically overestimated by this effect, as the triplet system has no Case A-Case B distinction since the $2^3S\rightarrow 1^1S$ transition is highly forbidden. Galaxy masses, star formation rates, or other properties that directly depend on the effective recombination coefficient of H$\alpha$ or H$\beta$ under the assumption of ``Case B'' would also be affected. The extinction coefficient $c(H\beta)$, computed with the H$\alpha$/H$\beta$ ratio, would also be affected, but to a relatively small degree that would only become noticeable in cases of significant deviations and high observational precision \citep{Scarlata:24}.

\subsection{$T_{\rm e}$(\oiii)$-$$T_{\rm e}$(\hei) and the Abundance Discrepancy Problem }
\label{sec:ADF}

The observation of $T_{\rm e}$(\hei) $<$ $T_{\rm e}$(\oiii) in a large sample of ionized nebulae aligns qualitatively with the temperature variations paradigm proposed by \citet{Peimbert:67}. In this paradigm, the presence of internal temperature variations in the gas introduces systematic biases toward higher temperatures in CEL diagnostics, as the emissivities of these lines have an exponential dependence on $T_{\rm e}$. This issue does not affect RL-based diagnostics, such as $T_{\rm e}$(\hei), which have a linear dependence on $T_{\rm e}$. An overestimation of $T_{\rm e}$ induced by CEL diagnostics will lead to a general underestimation of chemical abundances based on CEL ratios relative to H$\beta$ or H$\alpha$, which are RLs. Notably, in all nebular studies where heavy-element RLs have been detected, a systematic discrepancy has been found between determinations based on these lines and their collisionally excited counterparts \citep{Peimbert:03, Esteban:04, GarciaRojas:07, MendezDelgado:22b, Berg:24}. This longstanding problem, where CELs systematically yield lower abundances and the difference is usually quantified by the abundance discrepancy factor (ADF), dates back to the pioneering works of \citet{Bowen:39} and \citet{Wyse:42} and continues to be widely debated to this day.

\citet{Zhang:05} studied $T_{\rm e}$(\hei) and $T_{\rm e}$(\oiii) in a group of planetary nebulae using the same methodology as ours, described in Sec.~\ref{sec:obs}, with the difference that they used the effective recombination coefficients for \hei\ from \citet{Benjamin:99}. These authors found that $T_{\rm e}$(\hei) $<$ $T_{\rm e}$(\hi), which had been determined in previous works using the Balmer jump (BJ) and/or Paschen jump (PJ) continua. Among these works, the study by \citet{Liu:01} is particularly important, as they found that $T_{\rm e}$(\hi) $<$ $T_{\rm e}$(\oiii) and established a tight correlation between the ADF(O$^{2+}$) and $T_{\rm e}$(\oiii) - $T_{\rm e}$(\hi) (see their Fig.~8). Although the correlation between ADF(O$^{2+}$) and $T_{\rm e}$(\oiii) - $T_{\rm e}$(\hi) is one of the predictions of the \citet{Peimbert:67} formalism, the finding that $T_{\rm e}$(\hei) was lower than $T_{\rm e}$(\hi) was interpreted by \citet{Zhang:05} as evidence of a failure in the temperature fluctuation paradigm, which predicts $T_{\rm e}$(\hei) $\approx$ $T_{\rm e}$(\hi). 

\begin{figure}[h]
\epsscale{1.15}
\plotone{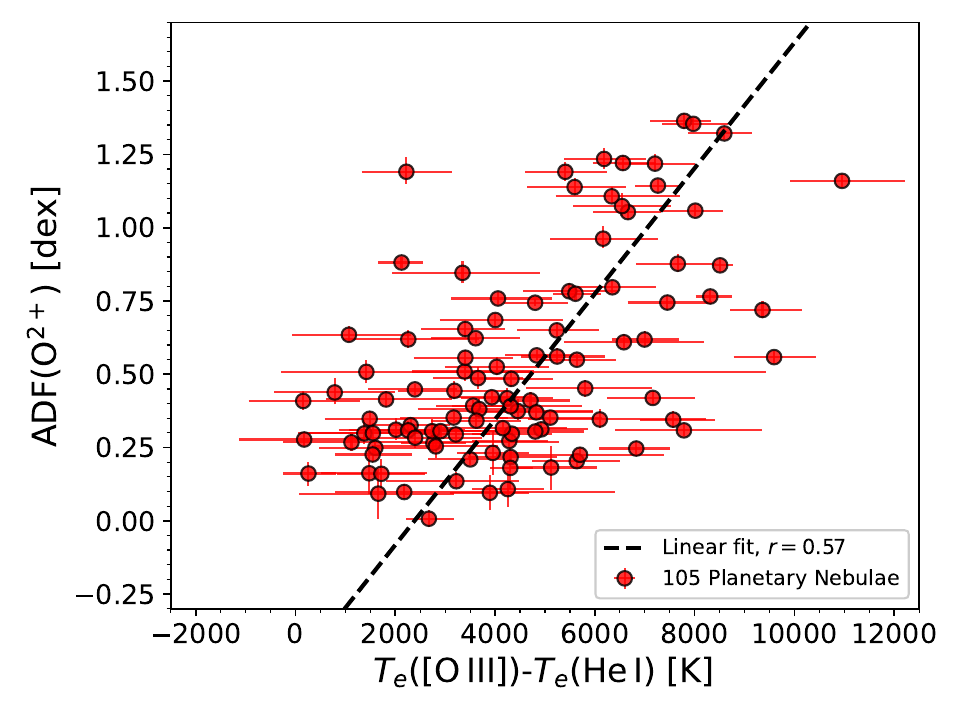}
\caption{Relationship between the ADF(O$^{2+}$) and the temperature difference $T_{\rm e}$(\oiii)$-$$T_{\rm e}$(\hei) for planetary nebulae, derived assuming the ``Case B'', considering objects where $T_{\rm e}$(\hei)$<$$T_{\rm e}$(\oiii). The ADF(O$^{2+}$) is the logarithmic difference between the O$^{2+}$/H$^{+}$ abundance determined using CELs (\oiii~$\lambda \lambda 4959, 5007$) and RLs (\oiirls~$\lambda \lambda 4638.86, 4641.81,$ $ 4649.13, 4650.84, 4661.63, 4673.73, 4676.23, 4696.35$), using the direct method and adopting $T_{\rm e}$(\oiii~$\lambda 4363/5007$). For consistency, only regions where the error in the flux of the \oiirls-RLs is equal to or less than 20\% were considered}
\label{fig:ADF_PNe}
\end{figure}

In Fig.~\ref{fig:ADF_PNe}, we show a correlation between $T_{\rm e}$(\oiii)$-$$T_{\rm e}$(\hei) and the ADF(O$^{2+}$) in planetary nebulae where $T_{\rm e}$(\hei)$<$$T_{\rm e}$(\oiii). We did not include ~\hii~regions or SFGs in the figure due to the limited number of these objects (23) with simultaneous determinations of ADF(O$^{2+}$) and the analyzed temperature difference. Additionally, the dynamic range of ADF(O$^{2+}$) covered by these specific regions is too limited to provide a comprehensive view of the abundance discrepancy problem in these objects, whose origin could be different \citep{MendezProceeding}. The trend observed in Fig.~\ref{fig:ADF_PNe} is consistent with the idea that hydrogen-poor clumps may exist in these objects, where cooling is especially efficient due to the concentration of heavy elements, as proposed by \citet{Torres-Peimbert:80} and \citet{Liu:01}. The linear fit of this relationship for PNe is shown in Eq.~\eqref{Eq:ADF_PNe}:
\begin{multline}
\label{Eq:ADF_PNe}
\text{ADF(O}^{2+}) = \left(2.15 \pm 0.23\right) \times 10^{-4}\\ \times \left[T_{\rm e}(\text{\oiii}) - T_{\rm e}(\text{\hei})\right]  
- (5.13 \pm 1.34) \times 10^{-1} ~[\text{dex}]
\end{multline}

However, it is important to note that under this paradigm, temperatures based on CEL ratios like $T_{\rm e}$(\oiii) may be overestimated due to auroral level population by recombination from high-metallicity clumps \citep{GomezLlanos:20,GomezLLanos:24,GarciaRojas:22}. In this hypothetical scenario, both O$^{2+}$ abundances based on CELs and those based on RLs using the direct method are conceptually incorrect, as this method assumes a homogeneous chemical composition. However, depending on the contrast between the different chemical components, sometimes the CELs and sometimes the RLs will be closer to the average abundances \citep{Morisset:23, MendezProceeding}.

Note, however, that Fig.~\ref{fig:ADF_PNe} is not inconsistent with the idea of temperature variations as proposed by \citet{Peimbert:67}. In contrast to the interpretation of \citet{Zhang:05}, the possibility that $T_{\rm e}$(\hei) $<$ $T_{\rm e}$(\hi) in PNe could be due to photon loss from the $n^1P \rightarrow 1^1S$ transitions, an effect not considered by \citet{Zhang:05}. The correlation between $T_{\rm e}$(\oiii)$-$$T_{\rm e}$(\hei) and the ADF(O$^{2+}$) in PNe shown in this work would not be lost when considering such effects, assuming the toy model described in Sec.~\ref{sec:toy_model}, as $T_{\rm e}(\text{\hei})$ and $T_{\rm e}(\text{\hei})_{TM}$ are linearly related, as shown in Fig.~\ref{Fig:toy_model_TheI_vs_caseB}. In any case, it would modify the slope and intercept shown in Eq.~\eqref{Eq:ADF_PNe}, but the correlation would remain. It is important to emphasize the need for deeper observations of PNe with low surface brightness to detect \oiirls-RLs under a wide range of physical conditions. This could help avoid hypothetical systematic biases toward particularly bright regions in \oiirls, which may not be representative of these objects in general.

For the case of \hii~regions and star-forming galaxies, if we follow the formalism of \citet{Peimbert:67}, assuming that $T_{\rm e}$(\hei) $\approx T_0$ \citep[see Eq.~11 in][]{Zhang:05} and considering Eq.~\eqref{Eq:eqtoymodel} with a typical temperature of $T_{\rm e}$(\oiii) $\approx 10,000$ K and Eq.~10 from \citet{Peimbert:13}, we obtain a root mean square deviation parameter ($t^2$) of $t^2 \approx 0.05$, which is typically found in star-forming regions to explain the ADF(O$^{2+}$) \citep{GarciaRojas:07b, Esteban:09, Peimbert:17, MendezDelgado:22b, Chen:24}.

The above discussion is consistent with the general finding that RLs seem to predict lower temperatures than CELs. This is true 
for both \oiirls~\citep{MendezDelgado:23a} 
and \cii~\citep{Torres-Peimbert:80}, and our work demonstrates that it could also be true for singlet-\hei. Additionally, \hi-RLs usually follow the same trend \citep[although with some contradictory examples, see][]{Guseva:07}. Recently, \citet{Khan:24} determined $T_{\rm e}$(\hi) using \hi~radio RLs and the radio continuum in 496 Galactic \hii~regions. The resulting temperature distribution predicts notably lower temperatures than those provided by CELs. For example, the regions M42, Sh~2-83, and M20 \citep{Esteban:04, Esteban:17, GarciaRojas:06}, located at Galactic distances of 8.54, 13.2, and 4.88 kpc \citep{MendezDelgado:22b}, have measured $T_{\rm e}$(\oiii) temperatures of 8370 K, 10370 K, and 7800 K, respectively, while the $T_{\rm e}$(\hi) predicted by the radio RLs is 7780 K, 9160 K, and 7270 K, respectively.

Our results highlight the need to understand the nature of the discrepancies between the physical conditions derived from CELs and RLs. If anomalies such as the ADF were to stem from RLs, it would imply fundamental errors in our understanding of recombination processes, potentially involving atoms like \hei~and \hi. To understand the ADF and nebular thermal structure, it is crucial to observe more faint and ultra-faint RLs, as these often reveal the most fundamental and important processes in our physical models. Relying solely on the analysis of strong CELs from different transitions (in UV, optical, or infrared wavelengths) seems insufficient, as it may not detect inconsistencies in the photoionization equilibrium processes, dominated by H and He, which emit exclusively RLs.

Regarding the tensions between CELs  and RLs, which sometimes exclusively focus on proving the absence or presence of temperature variations, it is important to address several points: (i) Potential shortcomings in the predictions of the $t^2$-formalism proposed by \citet{Peimbert:67} do not necessarily imply the absence of temperature variations; they may simply indicate that the modeling approach needs to be revised or it could indicate the existence of additional physical phenomena impacting the nebular abundances. Alternative paradigms also model temperature variations \citep[see the excellent discussion by][]{Stasinska:02}; (ii) The absence of temperature variations does not equate to the absence of problems. The existence of the ADF could have various causes. Depending on the source of this issue, the abundances derived from CELs might still be systematically incorrect \citep{MendezProceeding}.

\section{Summary and Final Thoughts}
\label{sec:concs}

In this manuscript, we report a systematic discrepancy between the observed fluxes of the singlet \hei~lines $\lambda \lambda 3614, 3965, 5016, 7281$ and the theoretical predictions under ``Case B'' in 393 optical spectra of Galactic and extragalactic \hii~regions, star-forming galaxies, and planetary nebulae. Of these spectra, 85 correspond to \hii~regions, 90 to star-forming galaxies, and the rest to planetary nebulae. In our analysis, we distinguish between star-forming regions and planetary nebulae. The observed \hei-singlet lines are systematically weaker than expected, as shown in Figs. \ref{Fig:distribution_7281_TO3}, \ref{Fig:distribution_5016_TO3}, \ref{Fig:distribution_3614_TO3}, and \ref{Fig:distribution_3965_TO3}. These discrepancies persist regardless of whether the recombination coefficients from \citet{Porter:12, Porter:13} or \citet{DelZanna:22} are used. The latter dataset significantly improves the consistency between theoretical predictions and observations for the \hei~$\lambda 6678$ and \hei~$\lambda 5876$ lines.

When determining $T_{\rm e}$(\hei) observationally using the \hei~$\lambda 7281/\lambda 5876$ and \hei~$\lambda 7281/\lambda 6678$ ratios under the ``Case B'' recombination model, we systematically find that $T_{\rm e}$(\hei) is lower than $T_{\rm e}$(\oiii~$\lambda 4363/\lambda 5007$) across most spectra for all types of objects. Additionally, we find that for the case of planetary nebulae where $T_{\rm e}$(\hei) $<$ $T_{\rm e}$(\oiii), there is a correlation between the discrepancy in abundances determined with \oiii-CELs and \oiirls-RLs, quantified by the ADF(O$^{2+}$), and $T_{\rm e}$(\oiii) $-$ $T_{\rm e}$(\hei). We have explored two potential explanations for the discrepancy between $T_{\rm e}$(\hei) and $T_{\rm e}$(\oiii): deviations from the ``Case B'' for $n^1P \rightarrow 1^1S$ \hei-transitions and the presence of temperature inhomogeneities, as proposed by \citet{Peimbert:67}. 

In the scenario where photons from the $n^1P \rightarrow 1^1S$ transitions are lost rather than being reabsorbed by \hei, we demonstrated that the toy model assuming a fraction of \hei~singlets are under ``Case B'' conditions while the rest are under ``Case A'' conditions fails to simultaneously explain the differences between the observations and the predictions for \hei~$\lambda \lambda 5016, 7281$ in all objets. However, in \hii~regions alone, this toy model is able to explain the observed discrepancies. For the rest of the objects, even if this model is operating in reality, an additional phenomenon, such as temperature variations, must be invoked. It is necessary to explore a ``Case C'' that includes fluorescent excitations $1^1S \rightarrow n^1P$ to shed more light on what is happening with the \hei~singlets. 

If we consider $T_{\rm e}(\text{\hei})_{TM}$ obtained from the toy model described earlier, this value remains systematically lower than $T_{\rm e}$(\oiii). The correlation between ADF(O$^{2+}$) and $T_{\rm e}$(\oiii) $-$ $T_{\rm e}$(\hei) in planetary nebulae is maintained when considering this toy model. The linear fit between $T_{\rm e}(\text{\hei})_{TM}$ and $T_{\rm e}$(\oiii) for \hii~regions and star-forming galaxies, represented in Eq.~\eqref{Eq:eqtoymodel}, is roughly consistent with what is expected under the presence of temperature inhomogeneities as proposed by \citet{Peimbert:67}. This linear fit predicts $t^2 \approx 0.05$ with $T_{\rm e}$(\oiii) $\approx 10,000$ K, a commonly observed case in the literature. However, this model is not self-consistent, as it assumes $T_{\rm e}$(\oiii) to determine the fraction of \hei~in ``Case B'' conditions. Considering only the ``Case B'' recombination, in consistency with Eq.~\eqref{Eq:TeHI_TeO3}, one may overestimate $t^2$ in certain regions of the $T_{\rm e}$(\oiii) parameter space.

If significant deviations from ``Case B'' are observed in the singlet system of \hei, it is important to understand why this issue is prevalent across a broad number of objects, which differ widely in their geometry, dust content, metallicity, and velocity gradients. We consider it possible that \hi~is absorbing some of the $n^1P \rightarrow 1^1S$ photons from the \hei~singlet system. However, this phenomenon does not seem to be naturally predicted by photoionization models, indicating radiative transfer effects that are generally not included. Such effects could also impact other elements with permitted transitions between different levels.

In a more speculative scenario, we also discuss the possibility that $n^1P \rightarrow 1^1S$ photons might be escaping into the ISM without being degraded into lower-energy photons. In that case, these photons could be contributing to the presence of the DIG, as they have energies around $\sim 22$ eV, capable of photoionizing a wide range of chemical elements. If this escape of ionizing photons occurs in the \hei~singlet system, it could also happen with \hi~due to their similar atomic configurations. This speculative case would affect many properties dependent on the emissivity of optical lines like H$\alpha$ or H$\beta$, such as chemical abundances, star formation rates, galaxy masses, and other quantities. Although speculative, this possibility should be interesting enough to test observationally whether there is a correlation between the observed deviations in \hei~singlets and the scape fraction of ionizing photons determined by other means.

In the case of temperature variations, one would need to explain the scale at which they operate and the physical phenomena that seem to generate them in most objects. The existence of high-metallicity clumps has been invoked in several studies to explain the low temperatures $T_{\rm e}$(\hei~$\lambda 7281/\lambda 6678$) found in planetary nebulae. While the creation and maintenance of such clumps in these objects remain subjects of debate, it seems a plausible explanation given the origin of planetary nebulae from the ejection of material from intermediate-mass stars in their late stages of life, which may not be well-mixed, especially in binary systems. However, extending this scenario to star-forming regions in a generalized manner seems implausible, as there is no accepted physical mechanism capable of generating large chemical variations within these objects. While generalized overheating from stellar feedback has been proposed in various studies to explain the ADF through temperature variations, no widely accepted physical mechanism consistent with the observations has been identified.

Our findings suggest that a combination of ionizing photon escape and temperature inhomogeneities may provide a better explanation for the observed discrepancies in \hei~$\lambda \lambda 5016, 7281$. Although other singlet lines such as \hei~$\lambda \lambda 3614, 3965$ also exhibit discrepancies in a similar manner to \hei~$\lambda \lambda 5016, 7281$, (See Figs \ref{Fig:distribution_3614_TO3} and \ref{Fig:distribution_3965_TO3}) their detections are considerably less frequent, and a deeper analysis of these lines is deferred to future work.  

\begin{acknowledgements}

JEMD and KK gratefully acknowledge funding from the Deutsche Forschungsgemeinschaft (DFG, German Research Foundation) in the form of an Emmy Noether Research Group (grant number KR4598/2-1, PI Kreckel) and the European Research Council’s starting grant ERC StG-101077573 (``ISM-METALS''). CE and JGR acknowledge financial support from the Agencia Estatal de Investigaci\'on of the Ministerio de Ciencia e Innovaci\'on (AEI- MCINN) under grant ``Espectroscop\'ia de campo integral de regiones H II locales. Modelos para el estudio de regiones H II extragal\'acticas'' with reference DOI:10.13039/501100011033. JGR also acknowledges financial support from the AEI-MCINN, under Severo Ochoa Centres of Excellence Programme 2020-2023 (CEX2019-000920-S), and from grant``Planetary nebulae as the key to understanding binary stellar evolution'' with reference PID-2022136653NA-I00 (DOI:10.13039/501100011033) funded by the Ministerio de Ciencia, Innovación y Universidades (MCIU/AEI) and by ERDF "A way of making Europe" of the European Union.

\end{acknowledgements}

\appendix
\setcounter{figure}{0}
\renewcommand{\thefigure}{A\arabic{figure}}

\section{Differences Between $T_{\rm e}$(\hei) and $T_{\rm e}$(\oiii) Arising from Ionization Differences in star forming regions}
\label{sec:appendix_temps}

To explore the natural deviations between $T_{\rm e}$(\hei)  and $T_{\rm e}$(\oiii) due to the slightly different ionization energies of He$^+$ (24.6-54.4 eV) and O$^{2+}$ (35.1-54.9 eV) in star forming regions, we explored photoionization models from the BOND project \citep{ValeAsari:16} using the Mexican Million Models database \citep{Morisset:15}. These models simulate giant ~\hii~regions under realistic ionization and metallicity conditions.

\begin{figure}
\epsscale{1.15}
\plotone{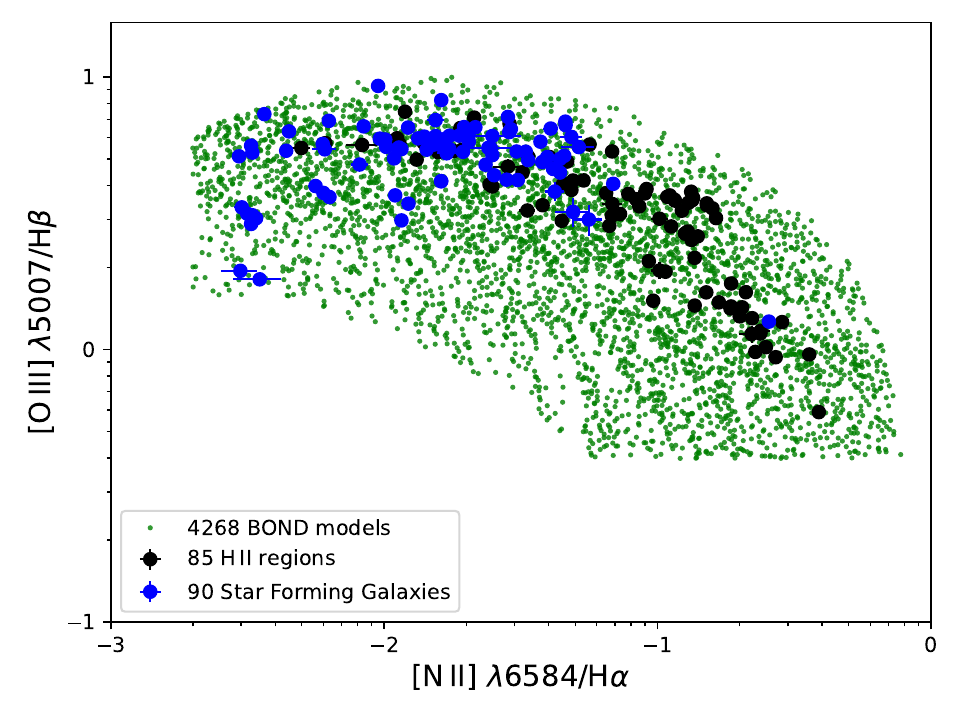}
\caption{Distribution of the selected photoionization models and the observed sample of \hii~regions and star-forming galaxies within the BPT diagram.}
\label{fig:MODEL_OBSER_BPT}
\end{figure}

We applied the selection criteria described in \citet{Amayo:21}, which were originally designed to ensure the consistency of photoionization model predictions with observed properties of \hii~regions. The only difference is that we adopted the following adjusted limits for the BPT diagram: log( \oiii/H$\beta$)$\geq -0.4$ and log( \nii/H$\alpha$)$\geq -2.7$, instead of the thresholds from \citet{Amayo:21} (log( \oiii/H$\beta$)$\geq -1.5$ and log( \nii/H$\alpha$)$\geq -2.5$). These adjustments ensured that our model sample better matches the parameter space of our observational data. As shown in Fig.~\ref{fig:MODEL_OBSER_BPT}, the observed star-forming regions are well represented within the BPT diagram by the selected models.

\begin{figure}
\epsscale{1.15}
\plotone{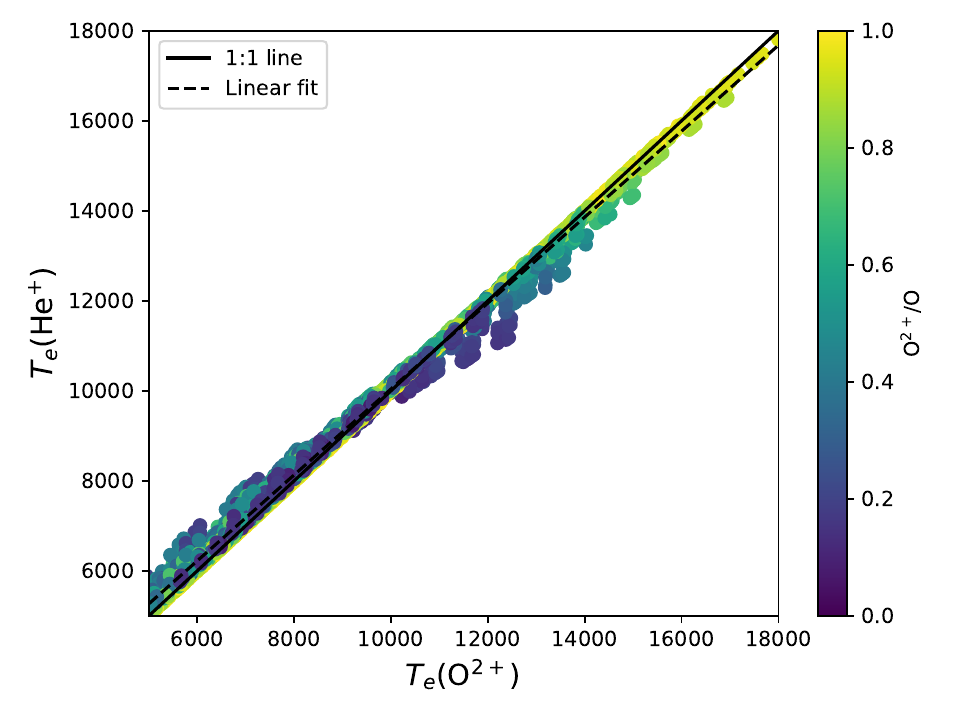}
\caption{Relationship between the temperatures $T_{\rm e}$(He$^{+}$) and $T_{\rm e}$(O$^{2+}$) predicted by the selected photoionization models. The color bar represents the ionization degree of the models.}
\label{fig:TeHeIIOIII}
\end{figure}

\begin{figure}
\epsscale{1.15}
\plotone{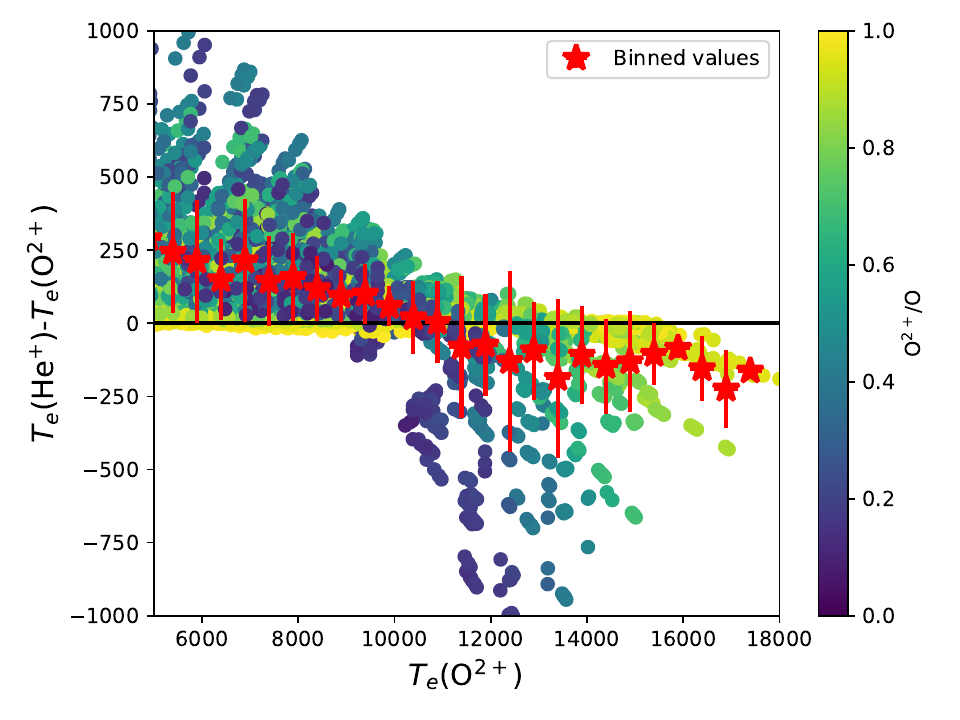}
\caption{Natural differences between $T_{\rm e}$(He$^{+}$) and $T_{\rm e}$(O$^{2+}$) predicted by the selected photoionization models.}
\label{fig:DiffTeHeIIOIII}
\end{figure}

From this selected sample, we analyzed the predicted electron temperatures for He$^+$ and O$^{2+}$ given by the photoionization models. The results are shown in Fig.~\ref{fig:TeHeIIOIII}. The linear fit of the data, represented by the dashed line is $T_{\rm e}$(He$^{+}$)=0.96$\times$$T_{\rm e}$(O$^{2+}$)+490, that closely aligns with the one-to-one line, with only small deviations. However, at $T_{\rm e}$(O$^{2+}$)$\lesssim 10 000$ K, some models show $T_{\rm e}$(He$^{+}$) $>$ $T_{\rm e}$(O$^{2+}$), whereas at higher temperatures, the opposite is observed. In order to quantify these difference, in Fig.~\ref{fig:DiffTeHeIIOIII}, we present the difference between $T_{\rm e}$(He$^{+}$) and $T_{\rm e}$(O$^{2+}$) as a function of $T_{\rm e}$(O$^{2+}$) , showing the differences in 500 K bins. As can be seen in the figure, the systematic differences are smaller than $\sim$250 K on average, significantly lower than those observed in Fig.~\ref{fig:TeHeI_vs_TOIII} derived from the observations.

\setcounter{figure}{0}
\setcounter{table}{0}
\renewcommand{\thefigure}{B\arabic{figure}}
\renewcommand{\thetable}{B\arabic{table}}
\section{Tables and figures}
\label{sec:appendix}

\begin{figure}
\epsscale{1.15}
\plotone{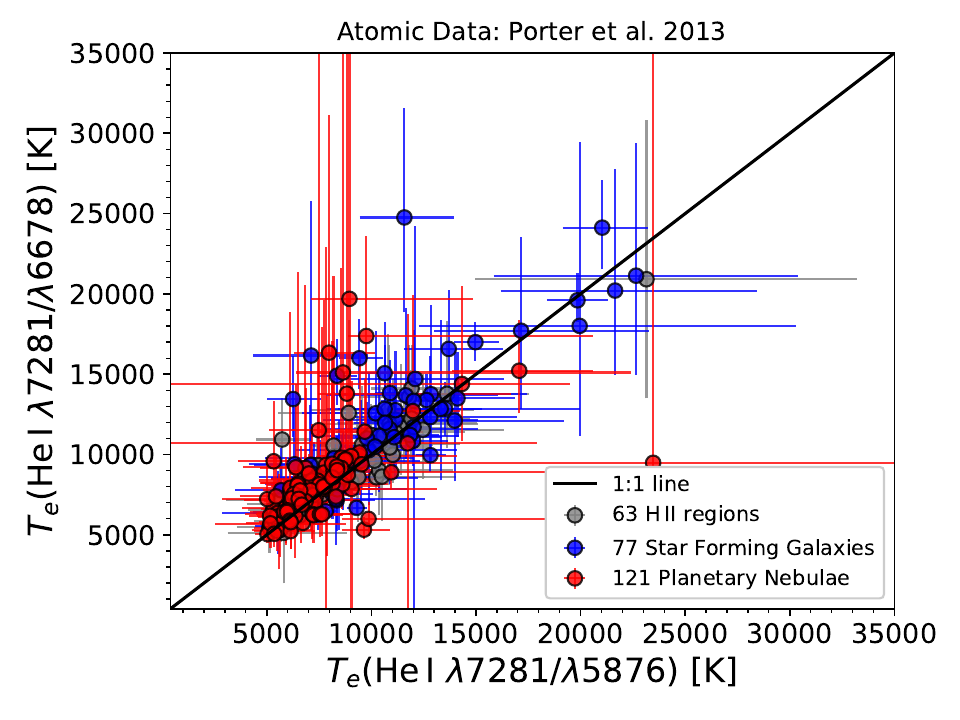}
\caption{Comparison between $T_{\rm e}$(\hei~$\lambda 7281/\lambda 5876$) and $T_{\rm e}$(\hei~$\lambda 7281/\lambda 6678$) determined with the atomic data from \citet{Porter:12, Porter:13}.}
\label{fig:TeHeI_6678_vs_THeI_5876_P1213}
\end{figure}

\begin{figure}
\epsscale{1.15}
\plotone{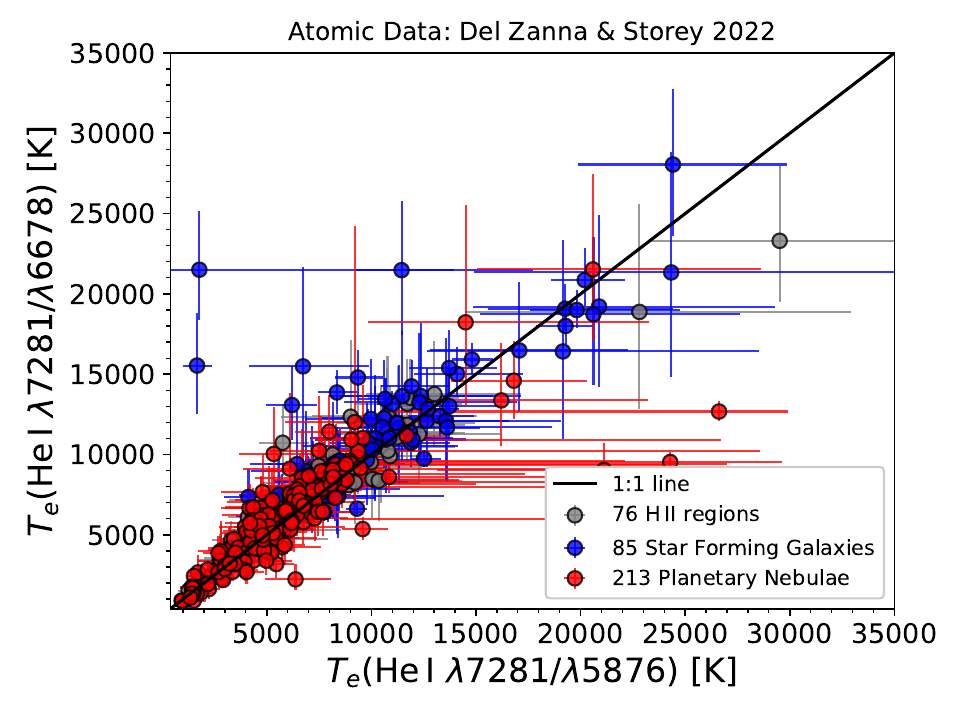}
\caption{The same as in Fig.~\ref{fig:TeHeI_6678_vs_THeI_5876_P1213} but considering the atomic data estimated by \citet{DelZanna:22}.}
\label{fig:TeHeI_6678_vs_THeI_5876_DZ22}
\end{figure}

\begin{figure}
\epsscale{1.15}
\plotone{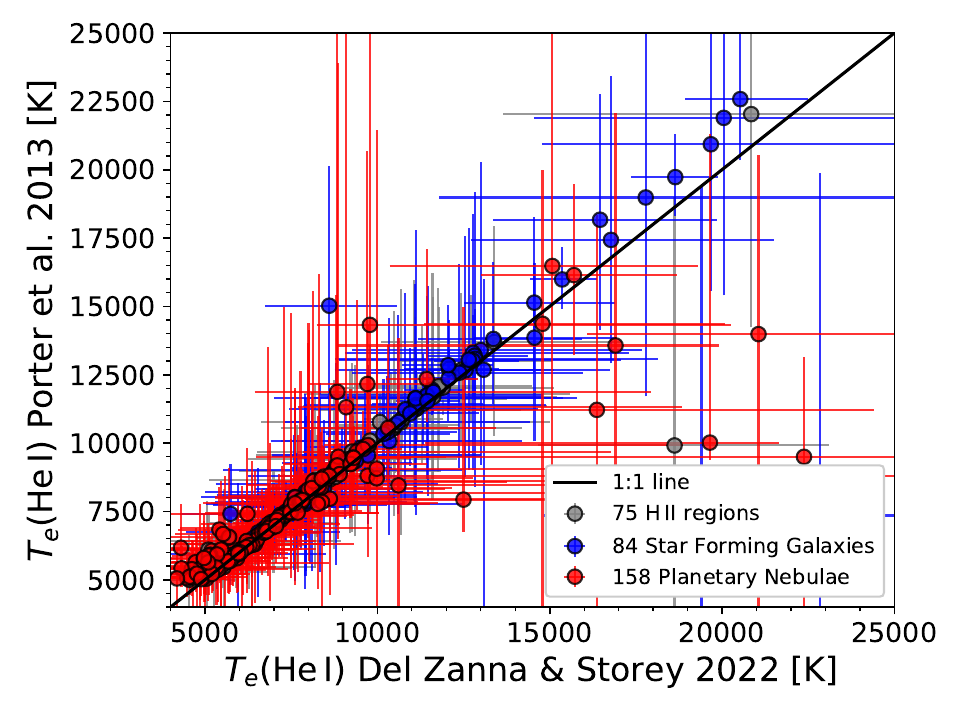}
\caption{Comparison between the final $T_{\rm e}$(\hei) values, resulting from averaging $T_{\rm e}$(\hei~$\lambda 7281/\lambda 5876$) and $T_{\rm e}$(\hei~$\lambda 7281/\lambda 6678$), obtained using the atomic data from \citet{Porter:12, Porter:13} and \citet{DelZanna:22}. The first set of atomic data defines the effective recombination coefficients for a temperature range between 5,000 and 25,000 K, whereas the second set of calculations covers a broader range, from 500 to 32,000 K}
\label{fig:general_Te_atomic_comparison}
\end{figure}

\clearpage

\begin{deluxetable*}{llll}
\tablecaption{Atomic data set used for collisionally excited lines\label{table:atomic_data}}
\tablewidth{0pt}
\tablehead{
Ion & Transition probabilities & Collision strengths}
\startdata
O$^{+}$   &  \citet{FroeseFischer:04} & \citet{Kisielius:09}\\
O$^{2+}$  &  \citet{Wiese:96}, \citet{Storey:00} & \citet{Aggarwal:99}\\
S$^{+}$   &  \citet{Irimia:05} & \citet{Tayal:10}\\
Cl$^{2+}$ &  \citet{Fritzsche:99} & \citet{Butler:89}\\
Ar$^{3+}$ &   \citet{Mendoza:82b}  & \citet{Ramsbottom:97}\\
Fe$^{2+}$ & \citet{Deb:09}, \citet{Mendoza:23} & \citet{Zhang:96}, \citet{Mendoza:23}\\
\enddata
\end{deluxetable*}

\startlongtable
\begin{deluxetable*}{cccccccc}
\tablecaption{Reference number, galaxy, region name and references to the sample of \hii~regions analyzed in this study \label{table:IDs_HII}}
\tablewidth{0pt}
\tablehead{
Reference number & Galaxy & Region &  Reference}
\startdata
HII-1 & LMC & 30Doradus &\citet{Peimbert:03}\\
HII-2 & LMC & IC2111 &\citet{DominguezGuzman:22}\\
HII-3 & LMC & N11B &\citet{DominguezGuzman:22}\\
HII-4 & LMC & N44C &\citet{DominguezGuzman:22}\\
HII-5 & LMC & NGC1714 &\citet{DominguezGuzman:22}\\
HII-6 & M101 & H1013 &\citet{Bresolin:07a}\\
HII-7 & M101 & H1013 &\citet{Croxall:16}\\
HII-8 & M101 & H1052 &\citet{Croxall:16}\\
\enddata
\end{deluxetable*}

\startlongtable
\begin{deluxetable*}{cccccccc}
\tablecaption{Reference number, galaxy, region name and references to the sample of Star Forming Galaxies analyzed in this study \label{table:IDs_SFG}}
\tablewidth{0pt}
\tablehead{
Reference number & Galaxy & Region &  Reference}
\startdata
SFG-1 & 0723+692A & - &\citet{Izotov:97b}\\
SFG-2 & 0930+554N & - &\citet{Izotov:97b}\\
SFG-3 & 1030+583 & - &\citet{Izotov:97b}\\
SFG-4 & 1054+365 & - &\citet{Izotov:97b}\\
SFG-5 & 1135+581 & - &\citet{Izotov:94}\\
SFG-6 & 1152+579 & - &\citet{Izotov:94}\\
SFG-7 & 1222+614 & - &\citet{Izotov:97b}\\
SFG-8 & 1223+487 & - &\citet{Izotov:97b}\\
\enddata
\end{deluxetable*}

\startlongtable
\begin{deluxetable*}{cccccccc}
\tablecaption{Reference number, galaxy, region name and references to the sample of planetary nebulae analyzed in this  \label{table:IDs_PNe}}
\tablewidth{0pt}
\tablehead{
Reference number & Galaxy & Region &  Reference}
\startdata
PNe-1 & LMC & SMP38 & Mampaso et al. (in prep)\\
PNe-2 & LMC & SMP52 &Mampaso et al. (in prep)\\
PNe-3 & LMC & SMP73 &Mampaso et al. (in prep)\\
PNe-4 & LMC & SMP92 &Mampaso et al. (in prep)\\
PNe-5 & MilkyWay & A12 &\citet{Henry:10}\\
PNe-6 & MilkyWay & BB1 &\citet{Kwitter:03}\\
PNe-7 & MilkyWay & Cn1-5 &\citet{GarciaRojas:12}\\
PNe-8 & MilkyWay & Cn2-1 &\citet{Wang:07}\\
\enddata
\end{deluxetable*}

\startlongtable
\begin{deluxetable*}{cccccccc}
\tablecaption{Electron densities derived from the sample of \hii~regions. \label{table:densities_HII}}
\tablewidth{0pt}
\tablehead{
Reference number& $n_{\rm e}$(\sii) & $n_{\rm e}$(\oii) & $n_{\rm e}$(\feiii) &$n_{\rm e}$(\cliii) &$n_{\rm e}$(\ariv) & Adopted $n_{\rm e}$  \\
& [cm$^{-3}$] & [cm$^{-3}$] & [cm$^{-3}$] & [cm$^{-3}$] & [cm$^{-3}$] & [cm$^{-3}$]}
\startdata
HII-1 & $340 \pm 20$ & $440 \pm 20$ & $1730 ^{+1840} _{-1670}$ & $330 ^{+260} _{-200}$ & $550 ^{+590} _{-300}$ & $390 \pm 20$ \\
HII-2 & $230 ^{+130} _{-110}$ & $250 ^{+120} _{-100}$ & $4600 ^{+9360} _{-3530}$ & $640 ^{+450} _{-440}$ & $2270 ^{+2270} _{-1390}$ & $240 ^{+130} _{-110}$ \\
HII-3 & $270 ^{+70} _{-60}$ & $270 ^{+120} _{-100}$ & $2860 ^{+2170} _{-2080}$ & $340 ^{+160} _{-150}$ & $680 ^{+610} _{-440}$ & $270 ^{+90} _{-80}$ \\
HII-4 & $110 \pm 50$ & $200 ^{+100} _{-90}$ & $2130 ^{+1870} _{-960}$ & $600 \pm 320$ & $580 \pm 370$ & $150 ^{+80} _{-70}$ \\
HII-5 & $360 ^{+140} _{-130}$ & $450 ^{+180} _{-140}$ & $4230 ^{+9660} _{-3160}$ & $620 ^{+500} _{-400}$ & $1580 ^{+1280} _{-1060}$ & $410 ^{+160} _{-130}$ \\
HII-6 & $30 ^{+50} _{-20}$ &  -  &  -  & $210 ^{+210} _{-130}$ &  -  & $100 \pm 100$ \\
HII-7 & $60 \pm 40$ & $220 ^{+40} _{-60}$ &  -  &  -  &  -  & $100 \pm 100$ \\
HII-8 & $140 \pm 60$ & $270 ^{+60} _{-50}$ &  -  &  -  &  -  & $210 \pm 60$ \\
\enddata
\end{deluxetable*}

\startlongtable
\begin{deluxetable*}{cccccccc}
\tablecaption{Electron densities derived from the sample of Star Forming Galaxies. \label{table:densities_SFG}}
\tablewidth{0pt}
\tablehead{
Reference number& $n_{\rm e}$(\sii) & $n_{\rm e}$(\oii) & $n_{\rm e}$(\feiii) &$n_{\rm e}$(\cliii) &$n_{\rm e}$(\ariv) & Adopted $n_{\rm e}$  \\
& [cm$^{-3}$] & [cm$^{-3}$] & [cm$^{-3}$] & [cm$^{-3}$] & [cm$^{-3}$] & [cm$^{-3}$]}
\startdata
SFG-1 & $80 ^{+60} _{-40}$ &  -  &  -  &  -  &  -  & $100 \pm 100$ \\
SFG-2 & $210 ^{+150} _{-120}$ &  -  &  -  &  -  &  -  & $210 ^{+150} _{-120}$ \\
SFG-3 & $20 \pm 10$ &  -  &  -  &  -  &  -  & $100 \pm 100$ \\
SFG-4 & $30 \pm 30$ &  -  &  -  &  -  &  -  & $100 \pm 100$ \\
SFG-5 & $30 \pm 20$ &  -  &  -  &  -  &  -  & $100 \pm 100$ \\
SFG-6 & $170 ^{+70} _{-80}$ &  -  &  -  &  -  &  -  & $170 ^{+70} _{-80}$ \\
SFG-7 & $40 ^{+30} _{-20}$ &  -  &  -  &  -  &  -  & $100 \pm 100$ \\
SFG-8 & $70 ^{+30} _{-40}$ &  -  &  -  &  -  &  -  & $100 \pm 100$ \\
\enddata
\end{deluxetable*}

\startlongtable
\begin{deluxetable*}{cccccccc}
\tablecaption{Electron densities derived from the sample of Planetary Nebulae. \label{table:densities_PNe}}
\tablewidth{0pt}
\tablehead{
Reference number& $n_{\rm e}$(\sii) & $n_{\rm e}$(\oii) & $n_{\rm e}$(\feiii) &$n_{\rm e}$(\cliii) &$n_{\rm e}$(\ariv) & Adopted $n_{\rm e}$  \\
& [cm$^{-3}$] & [cm$^{-3}$] & [cm$^{-3}$] & [cm$^{-3}$] & [cm$^{-3}$] & [cm$^{-3}$]}
\startdata
PNe-1 & $7890 ^{+4620} _{-2680}$ & $26330 ^{+38100} _{-10780}$ & $18260 ^{+68580} _{-10950}$ & $16350 ^{+6100} _{-4620}$ & $18850 ^{+1940} _{-1750}$ & $16640 \pm 4250$ \\
PNe-2 & $3780 ^{+1180} _{-770}$ & $5310 ^{+1400} _{-1030}$ &  -  & $4550 ^{+860} _{-760}$ &  -  & $4460 \pm 550$ \\
PNe-3 & $6770 ^{+3050} _{-1910}$ & $11930 ^{+6650} _{-3820}$ &  -  & $10960 ^{+1560} _{-1350}$ & $10920 ^{+970} _{-990}$ & $10560 \pm 1240$ \\
PNe-4 & $5690 ^{+1370} _{-1030}$ & $9360 ^{+4420} _{-2490}$ &  -  & $11940 ^{+1750} _{-1650}$ & $9390 ^{+930} _{-830}$ & $8690 \pm 2100$ \\
PNe-5 & $330 ^{+370} _{-210}$ &  -  &  -  &  -  &  -  & $330 ^{+370} _{-210}$ \\
PNe-6 &  -  &  -  &  -  &  -  &  -  & $100 \pm 100$ \\
PNe-7 & $3760 ^{+1890} _{-1070}$ & $5300 ^{+2530} _{-1530}$ & $12860 ^{+24750} _{-7860}$ & $4010 ^{+870} _{-900}$ & $10670 ^{+5410} _{-5820}$ & $4230 \pm 1010$ \\
PNe-8 & $4730 ^{+2440} _{-1410}$ & $7160 ^{+2880} _{-2110}$ &  -  & $9000 ^{+1840} _{-1270}$ & $19600 ^{+2610} _{-2480}$ & $9190 \pm 4780$ \\
\enddata
\end{deluxetable*}

\startlongtable
\begin{deluxetable*}{cccccccc}
\tablecaption{Electron temperatures derived from the sample of \hii~regions, considering recombination under ``Case B'' conditions. \label{table:temps_HII}}
\tablewidth{0pt}
\tablehead{
Reference number& $T_{\rm e}$(\oiii~$\lambda 4363/\lambda 5007$) & $T_{\rm e}$(\hei~$\lambda 7281/\lambda 5876$) & $T_{\rm e}$(\hei~$\lambda 7281/\lambda 6678$) & Adopted $T_{\rm e}$(\hei)  \\
& [K] & [K] & [K] & [K]}
\startdata
HII-1 & $9870 \pm 70$ & $4630 ^{+400} _{-410}$ & $4890 ^{+490} _{-440}$ & $4760 ^{+450} _{-420}$ \\
HII-2 & $9100 ^{+150} _{-170}$ & $4740 ^{+1000} _{-850}$ & $4920 ^{+1060} _{-820}$ & $4830 ^{+1030} _{-830}$ \\
HII-3 & $9150 ^{+90} _{-80}$ & $6880 ^{+580} _{-630}$ & $6740 ^{+640} _{-650}$ & $6810 ^{+610} _{-640}$ \\
HII-4 & $11300 ^{+160} _{-150}$ & $11900 ^{+1000} _{-990}$ &  -  & $11900 ^{+1000} _{-990}$ \\
HII-5 & $9520 ^{+210} _{-150}$ & $6790 ^{+1360} _{-1180}$ & $6510 ^{+1280} _{-1270}$ & $6650 ^{+1320} _{-1230}$ \\
HII-6 & $7640 ^{+210} _{-180}$ & $6060 ^{+1590} _{-1460}$ & $6370 ^{+1480} _{-1400}$ & $6210 ^{+1530} _{-1430}$ \\
HII-7 & $7440 ^{+170} _{-180}$ & $5720 ^{+680} _{-660}$ & $5760 ^{+830} _{-760}$ & $5740 ^{+750} _{-710}$ \\
HII-8 & $7730 ^{+90} _{-70}$ & $4320 ^{+580} _{-520}$ & $5540 ^{+810} _{-710}$ & $4930 ^{+700} _{-610}$ \\
\enddata
\end{deluxetable*}

\startlongtable
\begin{deluxetable*}{cccccccc}
\tablecaption{Electron temperatures derived from the sample of Star Forming Galaxies, considering recombination under ``Case B'' conditions. \label{table:temps_SFG}}
\tablewidth{0pt}
\tablehead{
Reference number& $T_{\rm e}$(\oiii~$\lambda 4363/\lambda 5007$) & $T_{\rm e}$(\hei~$\lambda 7281/\lambda 5876$) & $T_{\rm e}$(\hei~$\lambda 7281/\lambda 6678$) & Adopted $T_{\rm e}$(\hei)  \\
& [K] & [K] & [K] & [K]}
\startdata
SFG-1 & $15610 ^{+100} _{-90}$ & $11310 ^{+3870} _{-3490}$ & $11650 ^{+3930} _{-3460}$ & $11480 ^{+3900} _{-3470}$ \\
SFG-2 & $19990 ^{+330} _{-360}$ & $12360 ^{+4800} _{-3760}$ & $13640 ^{+4530} _{-3690}$ & $13000 ^{+4670} _{-3720}$ \\
SFG-3 & $15370 ^{+110} _{-180}$ & $17080 ^{+5200} _{-4270}$ & $16480 ^{+4250} _{-3860}$ & $16780 ^{+4730} _{-4070}$ \\
SFG-4 & $13720 ^{+180} _{-200}$ & $7170 ^{+3340} _{-2480}$ & $8040 ^{+3760} _{-2940}$ & $7600 ^{+3550} _{-2710}$ \\
SFG-5 & $13000 ^{+130} _{-120}$ & $6880 ^{+3090} _{-2630}$ & $8950 ^{+3730} _{-3180}$ & $7920 ^{+3410} _{-2900}$ \\
SFG-6 & $16110 ^{+190} _{-180}$ & $9970 ^{+3880} _{-3170}$ & $12220 ^{+3740} _{-3270}$ & $11100 ^{+3810} _{-3220}$ \\
SFG-7 & $14060 ^{+160} _{-110}$ & $12300 ^{+4680} _{-3610}$ & $13250 ^{+4340} _{-3790}$ & $12770 ^{+4510} _{-3700}$ \\
SFG-8 & $15980 ^{+100} _{-70}$ & $8360 ^{+3400} _{-3020}$ & $8290 ^{+3450} _{-3020}$ & $8330 ^{+3420} _{-3020}$ \\
\enddata
\end{deluxetable*}

\startlongtable
\begin{deluxetable*}{cccccccc}
\tablecaption{Electron temperatures derived from the sample of Planetary Nebulae, considering recombination under ``Case B'' conditions. \label{table:temps_PNe}}
\tablewidth{0pt}
\tablehead{
Reference number& $T_{\rm e}$(\oiii~$\lambda 4363/\lambda 5007$) & $T_{\rm e}$(\hei~$\lambda 7281/\lambda 5876$) & $T_{\rm e}$(\hei~$\lambda 7281/\lambda 6678$) & Adopted $T_{\rm e}$(\hei)  \\
& [K] & [K] & [K] & [K]}
\startdata
PNe-1 & $10920 ^{+140} _{-130}$ & $24290 ^{+5340} _{-15530}$ & $9530 ^{+650} _{-590}$ & $16910 ^{+3000} _{-8060}$ \\
PNe-2 & $13090 ^{+230} _{-180}$ & $8490 ^{+740} _{-650}$ & $9070 ^{+650} _{-680}$ & $8780 ^{+700} _{-670}$ \\
PNe-3 & $12310 ^{+170} _{-190}$ & $8070 ^{+580} _{-530}$ & $8650 ^{+530} _{-520}$ & $8360 ^{+550} _{-530}$ \\
PNe-4 & $13070 ^{+160} _{-180}$ & $8500 ^{+660} _{-580}$ & $9130 ^{+470} _{-490}$ & $8810 ^{+560} _{-540}$ \\
PNe-5 & $11480 ^{+640} _{-840}$ & $9200 ^{+4040} _{-3310}$ & $12020 ^{+12230} _{-5420}$ & $10610 ^{+8130} _{-4370}$ \\
PNe-6 & $13090 ^{+600} _{-860}$ & $9600 ^{+2950} _{-2480}$ & $11030 ^{+3310} _{-2590}$ & $10310 ^{+3130} _{-2540}$ \\
PNe-7 & $8770 ^{+140} _{-130}$ & $5310 ^{+1080} _{-880}$ & $6380 ^{+830} _{-980}$ & $5840 ^{+950} _{-930}$ \\
PNe-8 & $10100 ^{+200} _{-170}$ & $4840 ^{+590} _{-610}$ & $5940 ^{+590} _{-650}$ & $5390 ^{+590} _{-630}$ \\
\enddata
\end{deluxetable*}

\startlongtable
\begin{deluxetable*}{cccccccc}
\tablecaption{Abundances of O$^{2+}$/H$^{+}$ determined in the planetary nebulae sample and the ADF(O$^{2+}$). The abundances were determined using both \oiii-CELs and \oiirls-RLs, as well as the ``direct method'', assuming $T_{\rm e}$(\oiii) and adopted $n_{\rm e}$, as presented in Tables \ref{table:temps_PNe} and \ref{table:densities_PNe}, respectively. The abundances are presented in units of 12 + log(X/H). Only regions with a 20\% error or less in the flux of the \oiirls~RLs were considered. \label{table:ionic_abundances_PNe}}
\tablewidth{0pt}
\tablehead{
Reference number& O$^{2+}$ & O$^{2+}$ & ADF(O$^{2+}$) \\
& \oiii-CELs, $t^2 = 0$ & \oiirls-RLs &  }
\startdata
PNe-2 & $8.27 \pm 0.02$ & $8.49 \pm 0.04$ & $0.22 \pm 0.02$ \\
PNe-3 & $8.40 \pm 0.02$ & $8.63 ^{+0.08} _{-0.07}$ & $0.23 ^{+0.08} _{-0.07}$ \\
PNe-4 & $8.32 \pm 0.02$ & $8.43 \pm 0.06$ & $0.11 ^{+0.07} _{-0.06}$ \\
PNe-8 & $8.71 ^{+0.03} _{-0.02}$ & $9.11 \pm 0.03$ & $0.41 \pm 0.03$ \\
PNe-10 & $8.38 ^{+0.02} _{-0.03}$ & $8.73 ^{+0.06} _{-0.05}$ & $0.35 \pm 0.02$ \\
PNe-12 & $8.54 ^{+0.03} _{-0.02}$ & $9.16 \pm 0.03$ & $0.62 ^{+0.03} _{-0.02}$ \\
PNe-13 & $8.54 \pm 0.02$ & $8.96 \pm 0.03$ & $0.42 \pm 0.02$ \\
PNe-14 & $8.61 ^{+0.04} _{-0.03}$ & $9.03 \pm 0.07$ & $0.42 ^{+0.04} _{-0.03}$ \\
\enddata
\end{deluxetable*}

\startlongtable
\begin{deluxetable*}{cccccccc}
\tablecaption{Reddening-corrected \hei~fluxes in the sample of \hii~regions are reported by the reference authors. Intensities are given relative to H$\beta$ = 100. The equivalent width of H$\beta$ (EW(H$\beta$)) is also reported. \label{table:fluxes_HII}}
\tablewidth{0pt}
\tablehead{
Reference number & \hei~$\lambda 3614$ & \hei~$\lambda 3965$ & \hei~$\lambda 5016$ & \hei~$\lambda 5876$ &\hei~$\lambda 6678$ &\hei~$\lambda 7281$ & EW(H$\beta$)\\
& [H$\beta$=100] & [H$\beta$=100]& [H$\beta$=100]& [H$\beta$=100] & [H$\beta$=100] & [H$\beta$=100] & [\AA]  }
\startdata
HII-1 & $0.40 \pm 0.02$ & $0.86 \pm 0.03$ & $2.20 \pm 0.04$ & $12.1 \pm 0.1$ & $3.38 \pm 0.07$ & $0.45 \pm 0.02$ & $734 \pm 7$ \\
HII-2 &  -  & $0.89 \pm 0.05$ & $2.28 \pm 0.09$ & $11.8 \pm 0.6$ & $3.3 \pm 0.2$ & $0.44 \pm 0.03$ &  -  \\
HII-3 & $0.49 \pm 0.03$ & $0.78 \pm 0.05$ & $2.11 \pm 0.06$ & $11.5 \pm 0.2$ & $3.3 \pm 0.1$ & $0.52 \pm 0.02$ &  -  \\
HII-4 & $0.30 \pm 0.03$ & $0.60 \pm 0.04$ & $1.69 \pm 0.07$ & $9.8 \pm 0.2$ &  -  & $0.58 \pm 0.02$ &  -  \\
HII-5 & $0.51 \pm 0.08$ & $0.71 \pm 0.04$ & $2.22 \pm 0.09$ & $11.6 \pm 0.6$ & $3.4 \pm 0.2$ & $0.52 \pm 0.04$ &  -  \\
HII-6 &  -  &  -  & $2.4 \pm 0.1$ & $12.1 \pm 0.6$ & $3.4 \pm 0.2$ & $0.51 \pm 0.05$ & $146 \pm 6$ \\
HII-7 &  -  &  -  & $2.07 \pm 0.04$ & $11.6 \pm 0.2$ & $3.3 \pm 0.1$ & $0.47 \pm 0.02$ & $187 \pm 4$ \\
HII-8 &  -  & $1.1 \pm 0.1$ & $2.65 \pm 0.05$ & $14.0 \pm 0.3$ & $3.6 \pm 0.1$ & $0.51 \pm 0.03$ & $216 \pm 4$ \\
\enddata
\end{deluxetable*}

\startlongtable
\begin{deluxetable*}{cccccccc}
\tablecaption{Reddening-corrected \hei~fluxes in the sample of Star Forming Galaxies are reported by the reference authors. Intensities are given relative to H$\beta$ = 100. The equivalent width of H$\beta$ (EW(H$\beta$)) is also reported. \label{table:fluxes_SFG}}
\tablewidth{0pt}
\tablehead{
Reference number & \hei~$\lambda 3614$ & \hei~$\lambda 3965$ & \hei~$\lambda 5016$ & \hei~$\lambda 5876$ &\hei~$\lambda 6678$ &\hei~$\lambda 7281$ & EW(H$\beta$)\\
& [H$\beta$=100] & [H$\beta$=100]& [H$\beta$=100]& [H$\beta$=100] & [H$\beta$=100] & [H$\beta$=100] & [\AA]  }
\startdata
SFG-1 &  -  &  -  &  -  & $10.6 \pm 0.1$ & $2.90 \pm 0.09$ & $0.6 \pm 0.1$ & $357 \pm 4$ \\
SFG-2 &  -  &  -  &  -  & $9.9 \pm 0.3$ & $2.6 \pm 0.1$ & $0.6 \pm 0.1$ & $234 \pm 2$ \\
SFG-3 &  -  &  -  &  -  & $9.9 \pm 0.2$ & $2.8 \pm 0.1$ & $0.7 \pm 0.1$ & $92.0 \pm 0.9$ \\
SFG-4 &  -  &  -  &  -  & $10.9 \pm 0.3$ & $2.9 \pm 0.2$ & $0.5 \pm 0.1$ & $117 \pm 1$ \\
SFG-5 &  -  &  -  &  -  & $11.2 \pm 0.2$ & $2.8 \pm 0.1$ & $0.5 \pm 0.1$ & $128 \pm 1$ \\
SFG-6 &  -  &  -  &  -  & $11.1 \pm 0.2$ & $2.8 \pm 0.1$ & $0.6 \pm 0.1$ & $204 \pm 2$ \\
SFG-7 &  -  &  -  &  -  & $10.0 \pm 0.2$ & $2.7 \pm 0.1$ & $0.6 \pm 0.1$ & $87.0 \pm 0.9$ \\
SFG-8 &  -  &  -  &  -  & $10.2 \pm 0.1$ & $2.90 \pm 0.09$ & $0.5 \pm 0.1$ & $224 \pm 2$ \\
\enddata
\end{deluxetable*}

\startlongtable
\begin{deluxetable*}{cccccccc}
\tablecaption{Reddening-corrected \hei~fluxes in the sample of Planetary Nebulae are reported by the reference authors. Intensities are given relative to H$\beta$ = 100. \label{table:fluxes_PNe}}
\tablewidth{0pt}
\tablehead{
Reference number & \hei~$\lambda 3614$ & \hei~$\lambda 3965$ & \hei~$\lambda 5016$ & \hei~$\lambda 5876$ &\hei~$\lambda 6678$ &\hei~$\lambda 7281$ \\
& [H$\beta$=100] & [H$\beta$=100] & [H$\beta$=100] & [H$\beta$=100] & [H$\beta$=100] & [H$\beta$=100]  }
\startdata
PNe-1 &  -  & $0.83 \pm 0.03$ &  -  & $16.2 \pm 0.6$ & $4.3 \pm 0.2$ & $1.02 \pm 0.05$ \\
PNe-2 &  -  & $0.61 \pm 0.03$ &  -  & $11.6 \pm 0.3$ & $3.0 \pm 0.1$ & $0.64 \pm 0.03$ \\
PNe-3 &  -  & $0.42 \pm 0.02$ &  -  & $13.1 \pm 0.4$ & $3.4 \pm 0.1$ & $0.72 \pm 0.03$ \\
PNe-4 &  -  & $0.43 \pm 0.02$ & $1.22 \pm 0.06$ & $11.4 \pm 0.3$ & $2.94 \pm 0.09$ & $0.65 \pm 0.03$ \\
PNe-5 &  -  &  -  &  -  & $14 \pm 2$ &  -  & $0.7 \pm 0.1$ \\
PNe-6 &  -  &  -  &  -  & $11 \pm 1$ & $3.0 \pm 0.3$ & $0.60 \pm 0.06$ \\
PNe-7 & $0.53 \pm 0.07$ & $1.01 \pm 0.08$ & $2.9 \pm 0.3$ & $22 \pm 1$ & $5.7 \pm 0.3$ & $0.90 \pm 0.06$ \\
PNe-8 & $0.22 \pm 0.02$ &  -  &  -  & $18.2 \pm 0.9$ & $4.7 \pm 0.2$ & $0.71 \pm 0.04$ \\
\enddata
\end{deluxetable*}

\startlongtable
\begin{deluxetable*}{cccccccc}
\tablecaption{Predicted \hei~fluxes for the sample of \hii~regions considering $T_{\rm e}$(\oiii) and the adopted average $n_{\rm e}$ using the atomic data of \citet{DelZanna:22} under the ``Case B'' of recombination. The fluxes are normalized relative to \hei~$\lambda 5876$=100.\label{table:fluxes_predicted_HII}}
\tablewidth{0pt}
\tablehead{
Reference number & \hei~$\lambda 3614$ & \hei~$\lambda 3965$ & \hei~$\lambda 5016$ &\hei~$\lambda 6678$ &\hei~$\lambda 7281$ \\
& [\hei~$\lambda 5876$=100] & [\hei~$\lambda 5876$=100] & [\hei~$\lambda 5876$=100] & [\hei~$\lambda 5876$=100] & [\hei~$\lambda 5876$=100] }
\startdata
HII-1 & 4.06 & 8.28 & 20.85 & 28.16 & 5.45 \\
HII-2 & 3.96 & 8.10 & 20.39 & 28.36 & 5.16 \\
HII-3 & 3.97 & 8.11 & 20.41 & 28.34 & 5.18 \\
HII-4 & 4.27 & 8.70 & 21.78 & 28.19 & 5.74 \\
HII-5 & 4.01 & 8.19 & 20.64 & 28.23 & 5.33 \\
HII-6 & 3.74 & 7.65 & 19.35 & 28.54 & 4.69 \\
HII-7 & 3.70 & 7.57 & 19.17 & 28.55 & 4.62 \\
HII-8 & 3.75 & 7.68 & 19.42 & 28.50 & 4.75 \\
\enddata
\end{deluxetable*}

\startlongtable
\begin{deluxetable*}{cccccccc}
\tablecaption{Predicted \hei~fluxes for the sample of Star Forming Galaxies considering $T_{\rm e}$(\oiii) and the adopted average $n_{\rm e}$ using the atomic data of \citet{DelZanna:22} under the ``Case B'' of recombination. The fluxes are normalized relative to \hei~$\lambda 5876$=100.\label{table:fluxes_predicted_SFG}}
\tablewidth{0pt}
\tablehead{
Reference number & \hei~$\lambda 3614$ & \hei~$\lambda 3965$ & \hei~$\lambda 5016$ &\hei~$\lambda 6678$ &\hei~$\lambda 7281$ \\
& [\hei~$\lambda 5876$=100] & [\hei~$\lambda 5876$=100] & [\hei~$\lambda 5876$=100] & [\hei~$\lambda 5876$=100] & [\hei~$\lambda 5876$=100]}
\startdata
SFG-1 & 4.75 & 9.66 & 23.98 & 27.78 & 6.74 \\
SFG-2 & 4.86 & 9.90 & 24.57 & 26.17 & 7.61 \\
SFG-3 & 4.72 & 9.61 & 23.86 & 27.80 & 6.69 \\
SFG-4 & 4.56 & 9.28 & 23.10 & 28.00 & 6.31 \\
SFG-5 & 4.49 & 9.14 & 22.77 & 28.09 & 6.14 \\
SFG-6 & 4.71 & 9.59 & 23.84 & 27.31 & 6.90 \\
SFG-7 & 4.59 & 9.35 & 23.26 & 27.96 & 6.39 \\
SFG-8 & 4.78 & 9.73 & 24.13 & 27.73 & 6.82 \\
\enddata
\end{deluxetable*}

\startlongtable
\begin{deluxetable*}{cccccccc}
\tablecaption{Predicted \hei~fluxes for the sample of Planetary Nebulae considering $T_{\rm e}$(\oiii) and the adopted average $n_{\rm e}$ using the atomic data of \citet{DelZanna:22} under the ``Case B'' of recombination. The fluxes are normalized relative to \hei~$\lambda 5876$=100.\label{table:fluxes_predicted_PNe}}
\tablewidth{0pt}
\tablehead{
Reference number & \hei~$\lambda 3614$ & \hei~$\lambda 3965$ & \hei~$\lambda 5016$ &\hei~$\lambda 6678$ &\hei~$\lambda 7281$ \\
& [\hei~$\lambda 5876$=100] & [\hei~$\lambda 5876$=100] & [\hei~$\lambda 5876$=100] & [\hei~$\lambda 5876$=100] & [\hei~$\lambda 5876$=100]}
\startdata
PNe-1 & 3.70 & 7.59 & 19.75 & 25.37 & 6.80 \\
PNe-2 & 3.78 & 7.76 & 20.08 & 24.35 & 7.04 \\
PNe-3 & 3.68 & 7.55 & 19.70 & 24.30 & 7.15 \\
PNe-4 & 3.65 & 7.51 & 19.59 & 23.70 & 7.24 \\
PNe-5 & 4.24 & 8.65 & 21.71 & 27.90 & 5.87 \\
PNe-6 & 4.49 & 9.16 & 22.82 & 28.08 & 6.16 \\
PNe-7 & 3.77 & 7.71 & 19.72 & 27.50 & 5.62 \\
PNe-8 & 3.77 & 7.73 & 19.96 & 26.39 & 6.40 \\
\enddata
\end{deluxetable*}

\startlongtable
\begin{deluxetable*}{cccccccc}
\tablecaption{Predicted \hei~fluxes for the sample of \hii~regions considering $T_{\rm e}$(\hei) and the adopted average $n_{\rm e}$ using the atomic data of \citet{DelZanna:22} under the ``Case B'' of recombination. The fluxes are normalized relative to \hei~$\lambda 5876$=100.\label{table:fluxes_predicted_HII_THeI}}
\tablewidth{0pt}
\tablehead{
Reference number & \hei~$\lambda 3614$ & \hei~$\lambda 3965$ & \hei~$\lambda 5016$ &\hei~$\lambda 6678$ &\hei~$\lambda 7281$ \\
& [\hei~$\lambda 5876$=100] & [\hei~$\lambda 5876$=100] & [\hei~$\lambda 5876$=100] & [\hei~$\lambda 5876$=100] & [\hei~$\lambda 5876$=100] }
\startdata
HII-1 & 3.16 & 6.50 & 16.71 & 28.71 & 3.78 \\
HII-2 & 3.18 & 6.53 & 16.78 & 28.70 & 3.80 \\
HII-3 & 3.59 & 7.34 & 18.65 & 28.57 & 4.45 \\
HII-4 & 4.34 & 8.84 & 22.11 & 28.10 & 5.90 \\
HII-5 & 3.56 & 7.28 & 18.52 & 28.56 & 4.43 \\
HII-6 & 3.47 & 7.12 & 18.12 & 28.63 & 4.24 \\
HII-7 & 3.37 & 6.92 & 17.67 & 28.66 & 4.09 \\
HII-8 & 3.20 & 6.58 & 16.89 & 28.70 & 3.84 \\
\enddata
\end{deluxetable*}

\startlongtable
\begin{deluxetable*}{cccccccc}
\tablecaption{Predicted \hei~fluxes for the sample of Star Forming Galaxies considering $T_{\rm e}$(\hei) and the adopted average $n_{\rm e}$ using the atomic data of \citet{DelZanna:22} under the ``Case B'' of recombination. The fluxes are normalized relative to \hei~$\lambda 5876$=100.\label{table:fluxes_predicted_SFG_THeI}}
\tablewidth{0pt}
\tablehead{
Reference number & \hei~$\lambda 3614$ & \hei~$\lambda 3965$ & \hei~$\lambda 5016$ &\hei~$\lambda 6678$ &\hei~$\lambda 7281$ \\
& [\hei~$\lambda 5876$=100] & [\hei~$\lambda 5876$=100] & [\hei~$\lambda 5876$=100] & [\hei~$\lambda 5876$=100] & [\hei~$\lambda 5876$=100]}
\startdata
SFG-1 & 4.30 & 8.77 & 21.92 & 28.24 & 5.75 \\
SFG-2 & 4.43 & 9.02 & 22.54 & 27.76 & 6.22 \\
SFG-3 & 4.84 & 9.85 & 24.41 & 27.63 & 6.98 \\
SFG-4 & 3.73 & 7.64 & 19.31 & 28.54 & 4.67 \\
SFG-5 & 3.79 & 7.76 & 19.58 & 28.52 & 4.77 \\
SFG-6 & 4.24 & 8.65 & 21.65 & 28.19 & 5.69 \\
SFG-7 & 4.46 & 9.10 & 22.67 & 28.11 & 6.09 \\
SFG-8 & 3.86 & 7.88 & 19.87 & 28.49 & 4.89 \\
\enddata
\end{deluxetable*}

\startlongtable
\begin{deluxetable*}{cccccccc}
\tablecaption{Predicted \hei~fluxes for the sample of Planetary Nebulae considering $T_{\rm e}$(\hei) and the adopted average $n_{\rm e}$ using the atomic data of \citet{DelZanna:22} under the ``Case B'' of recombination. The fluxes are normalized relative to \hei~$\lambda 5876$=100.\label{table:fluxes_predicted_PNe_THeI}}
\tablewidth{0pt}
\tablehead{
Reference number & \hei~$\lambda 3614$ & \hei~$\lambda 3965$ & \hei~$\lambda 5016$ &\hei~$\lambda 6678$ &\hei~$\lambda 7281$ \\
& [\hei~$\lambda 5876$=100] & [\hei~$\lambda 5876$=100] & [\hei~$\lambda 5876$=100] & [\hei~$\lambda 5876$=100] & [\hei~$\lambda 5876$=100]}
\startdata
PNe-1 & 3.22 & 6.65 & 17.63 & 19.61 & 7.51 \\
PNe-2 & 3.76 & 7.70 & 19.71 & 27.47 & 5.63 \\
PNe-3 & 3.70 & 7.58 & 19.46 & 27.48 & 5.63 \\
PNe-4 & 3.72 & 7.63 & 19.61 & 27.22 & 5.81 \\
PNe-5 & 4.15 & 8.47 & 21.27 & 28.07 & 5.62 \\
PNe-6 & 4.15 & 8.47 & 21.24 & 28.34 & 5.44 \\
PNe-7 & 3.40 & 6.98 & 17.85 & 28.55 & 4.29 \\
PNe-8 & 3.32 & 6.82 & 17.48 & 28.60 & 4.15 \\
\enddata
\end{deluxetable*}

\startlongtable
\begin{deluxetable*}{cccccccc}
\tablecaption{Electron temperatures derived from the sample of \hii~regions, considering the toy model described in Sec.~\ref{sec:toy_model}, which accounts for mixed Case A-Case B conditions.\label{table:temps_HII_toy_model}}
\tablewidth{0pt}
\tablehead{
Reference number& $\gamma_{5016}$ & $T_{\rm e}$(\hei~$\lambda 7281/\lambda 5876$)$_{TM}$ & $T_{\rm e}$(\hei~$\lambda 7281/\lambda 6678$)$_{TM}$ & Average $T_{\rm e}$(\hei)$_{TM}$  \\
&  & [K] & [K] & [K]}
\startdata
HII-1 & 0.87 & $5390 \pm 490$ & $5670 ^{+550} _{-540}$ & $5530 ^{+550} _{-540}$ \\
HII-2 & 0.95 & $5010 ^{+1160} _{-850}$ & $5200 ^{+1050} _{-880}$ & $5110 ^{+1090} _{-870}$ \\
HII-3 & 0.89 & $7640 ^{+710} _{-570}$ & $7460 ^{+750} _{-620}$ & $7570 ^{+720} _{-620}$ \\
HII-4 & 0.78 & $14550 ^{+1230} _{-1190}$ &  -  & $14550 ^{+1230} _{-1190}$ \\
HII-5 & 0.93 & $7190 ^{+1300} _{-1130}$ & $6860 ^{+1320} _{-1200}$ & $7010 ^{+1310} _{-1160}$ \\
HII-7 & 0.93 & $6240 ^{+700} _{-720}$ & $6230 ^{+840} _{-770}$ & $6230 ^{+770} _{-750}$ \\
HII-8 & 0.97 & $4470 ^{+540} _{-550}$ & $5640 ^{+760} _{-700}$ & $4980 ^{+1030} _{-810}$ \\
HII-9 & 0.76 & $13560 ^{+4760} _{-3890}$ & $14850 ^{+4670} _{-4100}$ & $14200 ^{+4740} _{-4120}$ \\
\enddata
\end{deluxetable*}

\startlongtable
\begin{deluxetable*}{cccccccc}
\tablecaption{Electron temperatures derived from the sample of Star Forming Galaxies, considering the toy model described in Sec.~\ref{sec:toy_model}, which accounts for mixed Case A-Case B conditions. \label{table:temps_SFG_toy_model}}
\tablewidth{0pt}
\tablehead{
Reference number& $\gamma_{5016}$ & $T_{\rm e}$(\hei~$\lambda 7281/\lambda 5876$)$_{TM}$ & $T_{\rm e}$(\hei~$\lambda 7281/\lambda 6678$)$_{TM}$ & Average $T_{\rm e}$(\hei)$_{TM}$  \\
&  & [K] & [K] & [K]}
\startdata
SFG-14 & 0.98 & $7300 \pm 860$ & $7040 ^{+1330} _{-1060}$ & $7200 ^{+1080} _{-990}$ \\
SFG-15 & 0.65 & $9870 ^{+1220} _{-1110}$ & $10670 ^{+1310} _{-1120}$ & $10300 ^{+1330} _{-1250}$ \\
SFG-17 & 0.72 & $14610 ^{+1380} _{-1240}$ & $14820 ^{+760} _{-910}$ & $14740 ^{+980} _{-1120}$ \\
SFG-18 & 0.52 & $8690 ^{+810} _{-970}$ & $8910 ^{+770} _{-790}$ & $8810 ^{+780} _{-920}$ \\
SFG-19 & 0.35 & $14900 ^{+1690} _{-1490}$ & $14560 ^{+1520} _{-1290}$ & $14760 ^{+1580} _{-1410}$ \\
SFG-21 & 0.64 &  -  & $19140 ^{+2250} _{-1990}$ & $19140 ^{+2250} _{-1990}$ \\
SFG-22 & 0.73 & $11990 ^{+1630} _{-1280}$ & $18100 ^{+1840} _{-1880}$ & $14920 ^{+4070} _{-3520}$ \\
SFG-23 & 0.90 & $13820 ^{+770} _{-850}$ & $10680 ^{+500} _{-540}$ & $12070 ^{+2100} _{-1670}$ \\
\enddata
\end{deluxetable*}

\startlongtable
\begin{deluxetable*}{cccccccc}
\tablecaption{Electron temperatures derived from the sample of Planetary Nebulae, considering the toy model described in Sec.~\ref{sec:toy_model}, which accounts for mixed Case A-Case B conditions. \label{table:temps_PNe_toy_model}}
\tablewidth{0pt}
\tablehead{
Reference number& $\gamma_{5016}$ & $T_{\rm e}$(\hei~$\lambda 7281/\lambda 5876$)$_{TM}$ & $T_{\rm e}$(\hei~$\lambda 7281/\lambda 6678$)$_{TM}$ & Average $T_{\rm e}$(\hei)$_{TM}$  \\
&  & [K] & [K] & [K]}
\startdata
PNe-4 & 0.54 & $12930 ^{+3990} _{-1190}$ & $11800 ^{+590} _{-600}$ & $12140 ^{+1970} _{-780}$ \\
PNe-7 & 0.66 & $7250 ^{+1000} _{-870}$ & $7900 ^{+950} _{-820}$ & $7620 ^{+1000} _{-980}$ \\
PNe-10 & 0.95 & $7410 ^{+870} _{-780}$ & $8740 ^{+810} _{-790}$ & $8090 ^{+1090} _{-1060}$ \\
PNe-11 & 0.65 & $7750 ^{+950} _{-880}$ & $8240 ^{+880} _{-850}$ & $8000 ^{+970} _{-890}$ \\
PNe-14 & 0.61 & $8730 ^{+940} _{-860}$ & $8930 ^{+740} _{-690}$ & $8850 ^{+820} _{-780}$ \\
PNe-16 & 0.58 & $6880 ^{+1470} _{-1360}$ & $9570 ^{+1590} _{-1660}$ & $8210 ^{+2200} _{-2020}$ \\
PNe-22 & 0.93 & $11300 ^{+2230} _{-1810}$ & $9070 ^{+1130} _{-1140}$ & $9850 ^{+2380} _{-1530}$ \\
PNe-24 & 0.64 & $9840 ^{+1330} _{-870}$ & $9690 ^{+710} _{-700}$ & $9760 ^{+1010} _{-780}$ \\
\enddata
\end{deluxetable*}

\bibliography{sample631}{}
\bibliographystyle{aasjournal}



\end{document}